\documentclass[11pt]{article}
\usepackage{jheppub}
\usepackage{amsmath,amssymb,amsfonts,graphicx,slashed,color}
\usepackage{subcaption}
\usepackage{tikz}
\usepackage{tikz-3dplot}
\usepackage{pgfplots}
\usepgfplotslibrary{colormaps}
\usetikzlibrary{decorations.markings}
\usetikzlibrary{decorations.pathmorphing}
\usepackage[utf8]{inputenc}
\usepgfplotslibrary{fillbetween}
\usetikzlibrary{patterns}
\usetikzlibrary{shapes.misc}
\usepackage[utf8]{inputenc}
\usetikzlibrary{decorations.markings}
\usetikzlibrary{decorations.pathmorphing}
\newlength{\lx}
\newlength{\ly}
\newcommand{\be}{\begin{equation}}
\newcommand{\ee}{\end{equation}}
\newcommand{\bea}{\begin{eqnarray}}
\newcommand{\eea}{\end{eqnarray}}
\newcommand{\beq}{\begin{equation}}
\newcommand{\eeq}{\end{equation}}
\newcommand{\beqn}{\begin{eqnarray}}
\newcommand{\eeqn}{\end{eqnarray}}

\usepackage{upgreek}

\title{Soft photon theorems from CFT Ward identites in the flat limit of AdS/CFT}

\author[a]{Eliot Hijano,} \author[b]{ Dominik Neuenfeld}

\affiliation[\,a]{Department of Physics, Princeton University, Princeton, NJ 08544, USA.}
\affiliation[\,b]{Perimeter Institute for Theoretical Physics,
31 Caroline Street N., Waterloo, Ontario N2L 2Y5, Canada }

\emailAdd{hijano@princeton.edu}
\emailAdd{dneuenfeld@perimeterinstitute.ca}

\abstract{
S-matrix elements in flat space can be obtained from a large AdS-radius limit of certain CFT correlators. We present a method for constructing CFT operators which create incoming and outgoing scattering states in flat space. This is done by taking the flat limit of  bulk operator reconstruction techniques.  Using this method, we obtain explicit expressions for incoming and outgoing $U(1)$ gauge fields. Weinberg soft photon theorems then follow from Ward identites of conserved CFT currents. In four bulk dimensions, gauge fields on AdS can be quantized with standard and alternative boundary conditions. Changing the quantization scheme corresponds to the $S$-transformation of $SL(2, \mathbb Z)$ electric-magnetic duality in the bulk. This allows us to derive both, the electric and magnetic soft photon theorems in flat space from CFT physics.
}


\begin{document}

\tikzset{->-/.style={decoration={
  markings,
  mark=at position #1 with {\arrow{>}}},postaction={decorate}}}

\def \L {10}
\def \H {1.5*\L}

\tikzset{
    mark position/.style args={#1(#2)}{
        postaction={
            decorate,
            decoration={
                markings,
                mark=at position #1 with \coordinate (#2);
            }
        }
    }
}

\tikzset{
  pics/carc/.style args={#1:#2:#3}{
    code={
      \draw[pic actions] (#1:#3) arc(#1:#2:#3);
    }
  }
}

\tikzset{point/.style={insert path={ node[scale=2.5*sqrt(\pgflinewidth)]{.} }}}

\tikzset{->-/.style={decoration={
  markings,
  mark=at position #1 with {\arrow{>}}},postaction={decorate}}}

  \tikzset{-dot-/.style={decoration={
  markings,
  mark=at position #1 with {\fill[black] circle [radius=3pt,red];}},postaction={decorate}}} 
  
    \tikzset{-dotRed-/.style={decoration={
  markings,
  mark=at position #1 with {\fill[red] circle [radius=2pt,red];}},postaction={decorate}}} 
  
      \tikzset{-dotBlue-/.style={decoration={
  markings,
  mark=at position #1 with {\fill[blue] circle [radius=2pt,red];}},postaction={decorate}}} 
  
    \tikzset{-dotB-/.style={decoration={
  markings,
  mark=at position #1 with {\fill[black] circle [radius=3pt,red];}},postaction={decorate}}} 
  
    \tikzset{-dotW-/.style={decoration={
  markings,
  mark=at position #1 with {\fill[white] circle [radius=2pt,red];}},postaction={decorate}}} 

 \tikzset{-dot2-/.style={decoration={
  markings,
  mark=at position #1 with {\fill[blue] circle [radius=3pt,blue];}},postaction={decorate}}} 

    \definecolor{darkgreen}{RGB}{0,180,0}

    \definecolor{purple2}{RGB}{222,0,255}

 \tikzset{-dot3-/.style={decoration={
  markings,
  mark=at position #1 with {\fill[purple2] circle [radius=3pt,purple2];}},postaction={decorate}}} 

\tikzset{snake it/.style={decorate, decoration=snake}}

    \tikzset{cross/.style={cross out, draw=black, minimum size=2*(#1-\pgflinewidth), inner sep=0pt, outer sep=0pt},
cross/.default={3pt}}


\maketitle

\parskip=10pt
\addtocontents{toc}{\protect\setcounter{tocdepth}{2}}

\section{Introduction}
Despite much progress in the understanding of quantum gravity in asymptotically Anti-de Sitter spaces \cite{Maldacena:1997re, Witten:1998qj}, equally good insight into a holographic description of gravity in Minkowski space is still elusive. In this note we report on results which aim to further clarify the structure of holography in asymptotically flat space-times making use of the large radius limit of AdS/CFT.

In the presence of gravity, diffeomorphism invariance makes it impossible to define local correlation functions. While in AdS space, gauge invariant observables can be defined at the boundary, in the form of conformal correlators. Flat space has no such boundary. Instead, the central object and precise observable in flat space quantum gravity is the S-matrix. The reason is that at very early and late times, field excitations disperse and can be treated as essentially free.\footnote{This is not true in certain cases, most prominently if the particles live in four dimensions and couple to long range forces. In that case the particles do not decouple completely which gives rise to infrared divergences. However, several methods are known to extract finite scattering probabilities from IR divergent amplitudes.} In these asymptotic regions, all long range forces including gravity can be turned off (up to IR effects) and we can define the overlap between incoming and outgoing $n$-particle states, which defines S-matrix elements. Understanding flat space holography therefore means to have a prescription to calculate scattering amplitudes in a dual, lower-dimensional theory.   

In this paper, we will consider flat space as the large radius limit of Anti-de Sitter space. It was argued already very shortly after the discovery of the AdS/CFT correspondence that in this limit certain CFT correlation functions approach flat space S-matrix elements \cite{Polchinski:1999ry, Giddings:1999qu}. This has been elaborated upon in various works \cite{Giddings:1999jq,Gary:2009mi,Penedones:2010ue,Fitzpatrick:2011jn,Fitzpatrick:2011ia} and has found applications in the S-matrix bootstrap program \cite{Fitzpatrick:2011hu,Paulos:2016fap}.

The present paper has several goals. First, we give a more rigorous derivation of the CFT operators whose correlation functions turn into S-matrix elements of scalar particles. Our argument is an extension of previous work by one of the authors \cite{Hijano:2019qmi}. In \cite{Hijano:2019qmi} it was observed that CFT operators which create scattering states for massive and massless bulk fields can be derived in a unified fashion assuming HKLL bulk reconstruction \cite{Hamilton:2006az} in global AdS. The form of the CFT operators as the flat limit is taken then follows essentially from symmetry considerations. Here, we will again use bulk operator reconstruction, but take the flat space limit explicitly. This will give a formal derivation for flat space creation and annihilation operators in terms of CFT operators smeared along the time-like direction. For massless particles, the relevant contributions of those smeared operators are concentrated around a particular value of global time. In the massive case, operators must be smeared around complex points in the CFT. Acting with these operators on the vacuum creates scattering states. 

The method described above can be applied to fields with non-zero spin. As an example, we derive operators which create scattering states of $U(1)$ gauge bosons in terms of  smeared CFT operators. A gauge field on Anti-de Sitter space has multiple allowed boundary conditions and the expression for creation and annihilation operators depends on those. Quantization in AdS requires us to fix the boundary conditions of the magnetic or electric field at the asymptotic boundary, which we will refer to as ``magnetic'' and ``electric'' quantizations, respectively. For standard boundary conditions, which fix the magnetic field at the boundary, we find that bulk photon creation operators can be expressed as a particular smearing of a dual global $U(1)$ current. For example, for the case of the creation operator of an outgoing photon of negative helicity one finds
\begin{align}
\label{eq:IntroPhoton}
\sqrt{2\omega_{\vec{q}}} \,   a_{\vec{q}}^{\dagger(-)}=&   {-1\over 4\omega_{\vec{q}}}{1+z_q \bar{z}_q\over{\sqrt{2}}}  \int_0^{\pi} d\tau \,  e^{i \omega_{\vec{q}} L\left({\pi\over 2}-\tau \right)}
\int d^2 z {1\over (z_q-z)^2}j^+_{\bar{z}}(\tau,z,\bar{z}),
\end{align}
which takes the form of a $U(1)$ current smeared over the boundary. Here, $\omega_{\vec{q}}$ is the frequency of the photon and $(z_q,\bar{z}_q)$ specify the photon's direction. 
Conformal field theories associated to different boundary conditions in AdS are related by a $S$ transformation, which acts as electric-magnetic duality in the bulk \cite{Witten:2003ya}. If instead of fixing the magnetic field, we fix the electric field, the expression for creation/annihilation operators involves a CFT dynamical $U(1)$ gauge field. The expression equivalent to equation \eqref{eq:IntroPhoton} now involves a boundary gauge field inserted at a particular location on the boundary $S^2$ and smeared over time,
\begin{align}
\sqrt{2\omega_{\vec{q}}} \,   v_{\vec{q}}^{\dagger(-)}=&{1\over 4\omega_q}{1+z_q \bar{z}_q\over{\sqrt{2}}}  \int_0^{\pi} d\tau \,  e^{i \omega_{\vec{q}} L\left({\pi\over 2}-\tau \right)}
\partial_{\tau}A^+_{z}(\tau,z_q,\bar{z}_q).
\end{align}

These results allow to study how further properties of the S-matrix arise from CFT correlation functions. One such property are Weinberg's soft theorems \cite{Weinberg:1965nx}. Recently, a lot of work has been devoted to understanding them as Ward identities of asymptotic gauge transformations. For a review see \cite{Strominger:2017zoo}. Moreover, soft theorems for gravity have been proposed to play a role in resolutions to the black hole information paradox in flat space \cite{Hawking:2016msc}. Furthermore, it has been suggested that the asymptotic symmetry group of gravity in asymptotically flat spacetimes \cite{Sachs:1962zza,Bondi:1962px} plays an important role in the construction of a dual theory \cite{Strominger:2013jfa}.  In the present paper, we show that for the case of a $U(1)$ gauge theory, Weinberg soft theorems can be understood as arising from Ward identities of a CFT in the strict large $N$ limit.

The paper is organized as follows. In section \ref{sec:ScatteringStatesScalar} we give a formal derivation of CFT operators which create flat space scattering states. This will give a formula for scattering amplitudes of massive and massless scalars in terms of CFT correlation functions. Section \ref{sec:GaugeFields} discusses the two different quantization schemes for $U(1)$ gauge fields in $AdS_4$ and reviews how they are related to $SL(2, \mathbb Z)$ transformations of the dual CFT. We derive expressions for annihilation and creation operators for in and out-going photons for both quantization schemes. Section \ref{sec:WST} uses the results of previous sections to connect CFT Ward identities to Weinberg soft theorems. We close with a discussion in section \ref{sec:conc}.

\section{Scattering states revisited: Flat limit of AdS/CFT}\label{sec:ScatteringStatesScalar}
In this section we reconstruct scattering amplitudes of a theory in asymptotically flat space-times from the correlators of a conformal field theory in one lower dimension. This task has been the subject of much work in the literature \cite{Giddings:1999jq,Gary:2009mi,Giddings:1999qu,Balasubramanian:1999ri,Penedones:2010ue,Hijano:2019qmi}. Here we will revisit and re-phrase the argument in the language of path integrals, which are useful when discussing holography. We will focus on the case of bulk scalar fields, leaving the discussion of gauge theories for section \ref{sec:GaugeFields}.

\subsection{Scattering amplitudes in flat space-time}
\begin{figure}[]
\centering
\begin{tikzpicture}[scale=3]
\draw [white, thick] (-1,-1)--(1,1);
\draw [white, thick] (-1,1)--(1,-1);
\draw [black, thick] (-1,0)--(0,1);
\draw [black, thick] (1,0)--(0,1) node[pos=0.5,right=3]{${\cal I}^+$};
\draw [black, thick] (-1,0)--(0,-1);
\draw [black, thick] (1,0)--(0,-1) node[pos=0.5,right=3]{${\cal I}^-$};

\draw [red, very thick] (-1,0)   to[out=30,in=180]  (0,0.7) node [pos=1,above=9]{\,\,\scalebox{.4}{$\Sigma_+$}} node [pos=1,above=12,left=-1]{\,\,\scalebox{.4}{$| \text{out} \rangle$}};
\draw [red, very thick] (1,0)   to[out=150,in=0]  (0,0.7);

\draw [blue, very thick] (-1,0)   to[out=-30,in=180]  (0,-0.7) node [pos=1,below=9]{\,\,\scalebox{.4}{ $\Sigma_-$}} node [pos=1,below=10,right=1]{\,\,\scalebox{.4}{$| \text{in} \rangle$}};
\draw [blue, very thick] (1,0)   to[out=-150,in=0]  (0,-0.7);

\draw [gray, thick] (-1,0)   to[out=0,in=180]  (-0.5,0.1);
\draw [gray, thick] (0,0)   to[out=160,in=0]  (-0.5,0.1);
\draw [gray, thick] (0,0)   to[out=-20,in=180]  (0.5,-0.1);
\draw [gray, thick] (1,0)   to[out=180,in=0]  (0.5,-0.1);

\draw [black, thick,->] (-0.5,-0.3)--(-0.5,-0.1) node[pos=0.5,right]{$t$};
\end{tikzpicture}
\caption{Penrose diagram of flat space. Boundary conditions for bulk field operators are imposed at the early (late) time Cauchy slices $\Sigma_{\pm}$. The region of space-time before $\Sigma_-$ or after $\Sigma_+$ is characterized by an asymptotic Hamiltonian $H_{\text{as}}$ that we will consider to be free. 
}\label{fig:Mink}
\end{figure}
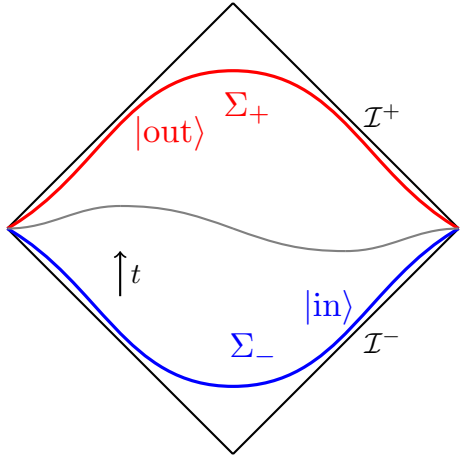

S-matrix elements of a quantum field theory in asymptotically flat space-time can be written as
\be\label{eq:SPI}
\langle \text{out}   \vert   \text{in}  \rangle = \sum_{\phi_{\text{a/b}}} \Psi^*_\text{out}[\phi_\text{b}] \int\displaylimits_{{\phi}\vert_{\Sigma_{-}}={\phi_\text{a}}}^{{\phi}\vert_{\Sigma_{+}}={\phi}_{\text{b}}} {\cal D}{\phi}\, e^{i S[{\phi}]} \Psi_\text{in}[\phi_\text{a}]\, ,
\ee
where ${\phi}$ stands for the fields of the theory with action $S[{\phi}]$, and the wave functionals $\Psi_\text{in}[\phi_\text{a}], \Psi^*_\text{out}[\phi_\text{b}]$ are given by overlaps of scattering states $| \text{out} \rangle, | \text{in} \rangle $ with field eigenstates $| \phi_\text{b,a} \rangle $ at very early and late Cauchy slices, which we will call $\Sigma_{\pm}$. Figure \ref{fig:Mink} gives a pictorial representation of formula \eqref{eq:SPI}.

In this paper, we consider standard Fock space scattering states $\vert\text{in}\rangle$ and $\langle \text{out}\vert$. The objective of scattering theory is to compute the overlap of two different such states. Ideally, one would like to consider scattering states which are states of the full Hamiltonian $H$ in the full Hilbert space ${\cal H}$ which asymptotically describe well separated particles. However, in interacting theories such states are usually unavailable and one is forced to simplify the problem by considering reasonable approximations to the full Hamiltonian at very early or very late times. Particles at those times are very far away from each other and can be understood as non-interacting, which is also referred to as \emph{asymptotically decoupled}. The asymptotic Hamiltonian $H_{\text{as}}$ approximating the dynamics in this case is free, and the asymptotic Hilbert space ${\cal H}_{as}$ takes the form of a Fock space.\footnote{To be more precise, one needs to identify vectors in the full Hilbert space that describe the states in the Fock space asymptotically. This amount to the construction of isometries implemented by M\o ller operators $\Omega_{\pm}$ which map states in ${\cal H}$ to states in ${\cal H}_{as}\subset {\cal H}$. The S-matrix in this language would then be $S=\Omega_-\Omega^{\dagger}_+$. For a more detailed discussion, see \cite{Buchholz:2005sa} and references therein.} \cite{Peskin, Weinberg}

In operator language, an incoming scattering state describing a single scalar particle with four-momentum $p$ corresponds to a field profile given by the following operator valued distribution
\be\label{eq:profile1P}
\lim_{t\rightarrow -\infty}{\phi}(t,\vec p) \sim {\phi}_{\text{in}}(t,\vec p) \sim \lim_{t\rightarrow -\infty} \left( e^{i t \omega_p}\hat{a}_\text{in}^{\dagger}(\vec{p}) + e^{- i \omega_p t}\hat{a}_\text{in}(\vec{p}) \right)\, ,
\ee
where  $\hat{a}_\text{in}(\vec{p})$ annihilates the vacuum. Acting with the in-field on the incoming vacuum creates the state $\vert \vec{p}\rangle \sim\hat{a}^{\dagger}(\vec{p}) \vert 0 \rangle$ at the Cauchy slice $\Sigma_-$ and it is a single particle state in the asymptotic Fock space. For out states, the story is analogous as $t \to \infty$. This demonstrates that scattering states are a very restricted class of states. In the language of path integrals, this places restrictions on the wave functionals defined in \eqref{eq:SPI}. In this and the next section, we will explain how such states can be prepared using a flat limit of AdS/CFT for fields of different spin. 

\subsection{Lorentzian AdS/CFT}
The question we want to answer is how the path integral in equation \eqref{eq:SPI} can be computed in the context of a flat limit of AdS/CFT. In order to do so, we first briefly review holography in AdS. We will be using global coordinates, for which the line element of empty AdS reads
\be\label{eq:ds2}
ds^2={L^2\over \cos^2\rho}\left(  -d\tau^2+d\rho^2+\sin^2\rho \, d\Omega_{d-1}^2   \right)\, .
\ee
The coordinate $\rho$ is holographic, in the sense that the boundary CFT is located at $\rho=\frac \pi 2$. The coordinates parametrising the CFT are $x=\{\tau,\Omega_{d-1}\}$, and $L$ denotes the AdS radius.

In order to point out a crucial difference between the two choices of signature, we first review Euclidean AdS holography. The AdS/CFT conjecture is an equality relating the generating functional of the correlators of the conformal field theory and a Euclidean path integral of bulk fields with specified boundary conditions. For a bulk field $\Phi$ with boundary condition
\be\label{eq:FallOffE}
\Phi \xrightarrow[\rho \to \frac \pi 2]{}  (\cos \rho)^{d-\Delta} \phi_E\, , 
\ee
the statement of the duality reads
\be\label{eq:EuclideanAdSCFT}
\langle e^{-\int_{\partial \text{AdS}} \phi_E {\cal O}} \rangle=\int [{\cal D}\Phi]_{\phi_E} \, e^{-S_E[\Phi]}\, ,
\ee
where $[{\cal D}\Phi]_{\phi_E}$ indicates that the path integral is over field configurations with asymptotic behavior \eqref{eq:FallOffE}. In the semi-classical limit, the bulk path integral can be computed in a saddle point approximation. Once the boundary condition $\phi_E$ is chosen, the saddle of the action is completely determined. The situation is different in Lorentzian signature \cite{Balasubramanian:1998de}, as conditions on the Lorentzian boundary do not fully specify a saddle of the action.
 
In the Lorentzian signature path integral, the bulk field $\Phi$ is integrated with a boundary condition $\phi_0(x)$ at the boundary. Such boundary condition corresponds to a source in the CFT, which couples to a dual operator ${\cal O}(x)$ with conformal dimension $\Delta$. In a semi-classical approximation, and unlike in Euclidean signature, the saddles of the action are however not completely determined by the boundary condition. The general structure of the field solution reads
\be
\Phi=\tilde{\phi}(\rho,x)+{\phi}(\rho,x)\, ,
\ee
where we have split the field into pieces with different fall-off at the AdS boundary
\be
\tilde{\phi}(\rho,x) \xrightarrow[\rho \to \frac \pi 2]{} (\cos\rho)^{d-\Delta} \phi_0(x)\, , \quad \text{and}\quad {\phi}(\rho,x) \xrightarrow[\rho \to \frac \pi 2]{} (\cos\rho)^{\Delta} {\phi}(x)\, .
\ee
The part of $\Phi$ that asymptotes to $\phi_0$ is commonly known as the non-normalizable part of the field, while the sub-leading structure  ${\phi}(\rho,x)$  is known as the normalizable part and captures the remaining degrees of freedom not fixed by the boundary condition\footnote{
Even though we will refer to the source part of the bulk field as the non-normalizable modes and the state part of the field as the normalizable modes, for the range of bulk mass $-d^2/4<m^2<1-d^2/4$, both sets of modes are normalizable, and some criterion is needed to distinguish them \cite{Balasubramanian:1998sn}. We will revisit these statements more carefully in section \ref{sec:ScalarLagrange}
}. The modes of normalizable solutions are low energy excitations of space-time and can be thought of as spanning the low energy Hilbert space of the boundary theory. At the microscopic level, we regard $\hat{{\phi}}(\rho,x)$ as a quantized operator that will match ${\cal O}(x)$ at the boundary as
\be
\hat{{\phi}}(\rho,x) \xrightarrow[\rho \to \frac \pi 2]{} (\cos\rho)^{\Delta}{\cal O}(x)\, .
\ee
Note that here $\mathcal O(x)$ is the CFT operator and not just its expectation value.
The fact that the normalizable field is not completely determined by the boundary conditions allows us to specify an incoming and an outgoing state in AdS by specifying a scattering state at some Cauchy slices $\Sigma_{\pm}^{\text{AdS}}$ in the bulk. This in turn also specifies a particular state in the CFT. Throughout this paper, one of the issues we will discuss is precisely how to construct CFT states $\langle \text{in} \vert$ and $\vert \text{out} \rangle$ so that
\be\label{eq:NormPhiZ}
\langle \text{in} \vert   e^{i \int_{\partial \text{AdS}} \phi_0(x){\cal O}(x)} \vert \text{out}\rangle = \sum_{\phi_{\text{a/b}}} \Psi^*_\text{out}[\phi_\text{b}]  \int\displaylimits_{{\phi}\vert_{\Sigma^{\text{AdS}}_{-}}={\phi}_{\text{a}}}^{{\phi}\vert_{\Sigma^{\text{AdS}}_{+}}={\phi}_{\text{b}}} [{\cal D}\Phi]_{\phi_0}\, e^{i S({\phi})} \Psi_\text{in}[\phi_\text{a}]\, .
\ee
Here, $\phi_0$ is the the non-normalizable part of the field, and field configurations ${\phi}_{\text{a/b}}$ fix the normalizable part of the field at Cauchy slices $\Sigma^{\text{AdS}}_{\pm}$.

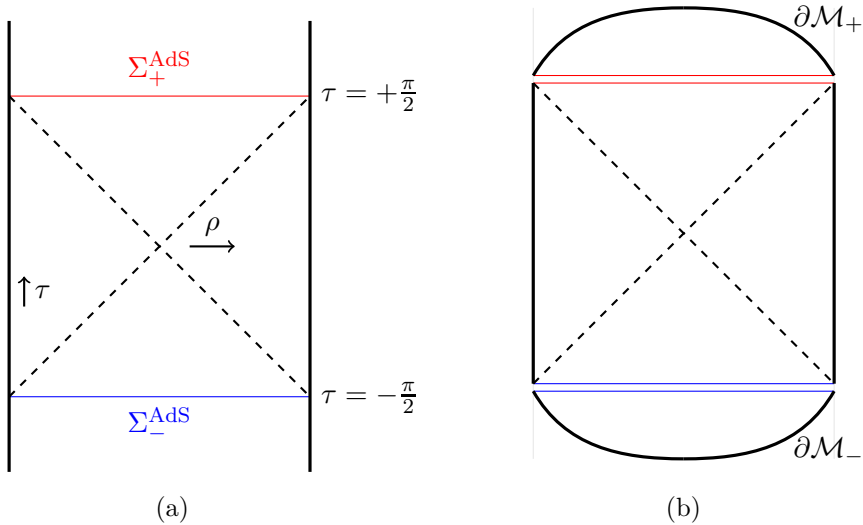
\begin{figure}[]
\centering
\begin{tabular}{cc}
\begin{subfigure}[t]{0.41\textwidth}
\centering
\begin{tikzpicture}[scale=1]
\draw [white, thick] (-3,-3)--(3,3);
\draw [white, thick] (-3,3)--(3,-3);

\draw[red](-2,2)--(2,2) node[pos=0.5,above]{$\Sigma_+^{\text{AdS}}$};
\draw[blue](-2,-2)--(2,-2) node[pos=0.5,below]{$\Sigma_-^{\text{AdS}}$};

\draw [black, thick,dashed] (-2,-2)--(2,2);
\draw [black, thick,dashed] (-2,2)--(2,-2);

\draw [black, very thick](-2,-3)--(-2,3);
\draw [black, very thick](2,-3)--(2,3) node[pos=0.165, right]{$\tau=-{\pi \over 2}$} node[pos=0.835, right]{$\tau=+{\pi \over 2}$};
\draw [black, thick,->] (-1.8,-0.8)--(-1.8,-0.4) node[pos=0.5,right]{$\tau$};
\draw [black, thick,->] (0.4,0)--(1.0,0) node[pos=0.5,above]{$\rho$};
\end{tikzpicture}
\caption{}
\end{subfigure}
&
\begin{subfigure}[t]{0.41\textwidth}
\centering
\begin{tikzpicture}[scale=1]
\draw [white, thick] (-3,-3)--(3,3);
\draw [white, thick] (-3,3)--(3,-3);

\draw[red](-2,2)--(2,2) ;
\draw[blue](-2,-2)--(2,-2);

\draw[red](-2,2.1)--(2,2.1) ;
\draw[blue](-2,-2.1)--(2,-2.1);

\draw [black, thick,dashed] (-2,-2)--(2,2);
\draw [black, thick,dashed] (-2,2)--(2,-2);

\draw [gray,opacity=0.2](-2,-3)--(-2,3);
\draw [gray,opacity=0.2](2,-3)--(2,3);
\draw [black, very thick](-2,-2)--(-2,2);
\draw [black, very thick](2,-2)--(2,2) node[pos=0,left=2,below=16]{$\partial{\cal M}_-$} node[pos=1,left=2,above=15]{$\partial{\cal M}_+$};
\draw [black, very thick] (-2,-2.1)   to[out=-60,in=-180]  (0,-3);
\draw [black, very thick] (2,-2.1)   to[out=-120,in=0]  (0,-3);
\draw [black, very thick] (-2,2.1)   to[out=60,in=-180]  (0,3);
\draw [black, very thick] (2,2.1)   to[out=120,in=0]  (0,3);
\end{tikzpicture}
\caption{}
\end{subfigure}
\end{tabular}
    \caption{ a) Penrose diagram of Anti-de Sitter space. Boundary conditions for normalizable bulk field operators can be imposed at the Cauchy slices $\Sigma_{\pm}^{\text{AdS}}$. The bulk path integral computes \eqref{eq:NormPhiZ}. b) Gluing Lorentzian AdS to Euclidean half-spheres ${\cal M}_{\pm}$ and placing boundary conditions for the bulk field at ${\partial}{\cal M}_{\pm}$ yields the SvR prescription. This corresponds to a CFT expectation value involving coherent states.
}\label{fig:AdS}
\end{figure}

Before we do so, let us contrast our approach to an existing way to construct CFT excited states in the context of real time AdS/CFT; The Skenderis-van Rees framework (SvR) \cite{Skenderis:2008dh,Skenderis:2008dg}. In their formalism, the states of the CFT are prepared using a Euclidean path integral at the boundary, which can be shown to correspond to preparing coherent states in the AdS Fock space \cite{Botta-Cantcheff:2015sav}. For a picture, see figure \ref{fig:AdS}. Their construction takes the form \eqref{eq:NormPhiZ}.
In the case of initial states defined on $\Sigma^{\text{AdS}}_{-}$, we glue the Lorentzian space-time at $\Sigma_{-}^{\text{AdS}}$ to a Euclidean manifold ${\cal M}_-$. The state is then specified by a boundary condition of the bulk field at $\partial{\cal M}_-$. The wave functional which defines the incoming state thus reads
\be
\Psi_\text{in}[{\phi}_{\text{a}}]\equiv \langle {\phi}_{\text{a}}\vert \Psi_\text{in} \rangle = \int_0^{{\phi}_{\text{a}}} [{\cal D}\Phi]_{\phi_i} e^{-S_{{\cal M}_-}[\Phi]}\, .
\ee
Here, $\phi_i$ is a boundary condition for the non-normalizable part of the field at the Euclidean boundary $\partial {\cal M}_-$. At Euclidean infinite past, the field is set to zero, while  ${\phi}_{\text{in}}$ is the boundary condition for the field at $\Sigma_-^{\text{AdS}}$. The CFT state dual to this prescription corresponds to \cite{Botta-Cantcheff:2015sav}
\be
\vert \Psi_i \rangle = e^{-\int_{\partial {\cal M}_-} {\cal O}\phi_i}\vert 0 \rangle\, .
\ee
Expanding ${\cal O}$ in creation and annihilation operators shows that $\vert \Psi_i\rangle$ is a coherent state. Final states are obtained by a similar construction using a Euclidean path integral over ${\cal M}_+$ and glued onto $\Sigma^{\text{AdS}}_{+}$.

There are two main differences between our approach and that of SvR. First, we will use an alternative procedure that does not rely on Euclidean path integrals. Instead we will construct states $\vert \text{in} \rangle$ (and $\langle \text{out} \vert$) by acting with certain bulk operators (in Lorentzian signature) on the vacuum of the theory. Using a bulk operator reconstruction method, we will be able to construct such states as operators in the CFT acting on the vacuum. Second, the SvR framework is quite general and allows to discuss all states we can prepare as a Euclidean path integral. In contrast, we will be interested in CFT operators, which create bulk states that turn into Fock space scattering states in the large radius limit.

\subsection{Flat limit of AdS/CFT}
\begin{figure}[]
\centering
\begin{tikzpicture}[scale=1.2]
\draw [white, thick] (-2,-2)--(2,2);
\draw [white, thick] (-2,2)--(2,-2);
\draw [black, thick] (-1,0)--(0,1);
\draw [black, thick] (1,0)--(0,1);
\draw [black, thick] (-1,0)--(0,-1);
\draw [black, thick] (1,0)--(0,-1);

\draw [red, very thick] (-1,0)   to[out=30,in=180]  (0,0.7);
\draw [red, very thick] (1,0)   to[out=150,in=0]  (0,0.7);

\draw [blue, very thick] (-1,0)   to[out=-30,in=180]  (0,-0.7);
\draw [blue, very thick] (1,0)   to[out=-150,in=0]  (0,-0.7);

\draw [red, very thick] (-2,0.01)--(-1,0.01) node [pos=0.5,above=0]{\,\,{ $\Sigma^{\prime}_+$}};
\draw [blue, very thick] (-2,-0.01)--(-1,-0.01);

\draw [red, very thick] (2,0.01)--(1,0.01);
\draw [blue, very thick] (2,-0.01)--(1,-0.01)node [pos=0.5,below=0]{\,\,{ $\Sigma^{\prime}_-$}};

\draw[red](-2,2)--(2,2) node[midway,below]{$\Sigma^{\text{AdS}}_+$} ;
\draw[blue](-2,-2)--(2,-2)  node[midway,above]{$\Sigma^{\text{AdS}}_-$};

\draw [black, very thick](-2,-2)--(-2,2);
\draw [black, very thick](2,-2)--(2,2);

\draw [black, thick,->] (-1.8,-1.3)--(-1.8,-0.9) node[pos=0.5,right]{$\tau$};
\draw [black, thick,->] (1.1,-1.3)--(1.7,-1.3) node[pos=0.5,above]{$\rho$};

\end{tikzpicture}
    \caption{Section of the global AdS cylinder. The region inside the diamond corresponds to $\tau=t/L$ and $\rho=r/ L$ which turns flat as $L$ is taken to infinity.  The Cauchy slices $\Sigma_{\pm}^{\prime}$ extend the Minkowski slices $\Sigma_{\pm}$ to global AdS space. Specifying a state on $\Sigma_{\pm}^{\prime}$ specifies a state in the conformal field theory, in the Cauchy slice $\Sigma_{\pm}^{\prime}\cap \partial\text{AdS}$. Alternatively, we can invoke asymptotic decoupling and use a free Hamiltonian in AdS to time evolve our fields to the Cauchy slices $\Sigma_{\pm}^{\text{AdS}}$. This would yield the computation of formula \eqref{eq:NormPhiZ} with vanishing source ($\phi_0=0$), which is drawn in figure \ref{fig:AdS} a).
}\label{fig:Limit}
\end{figure}
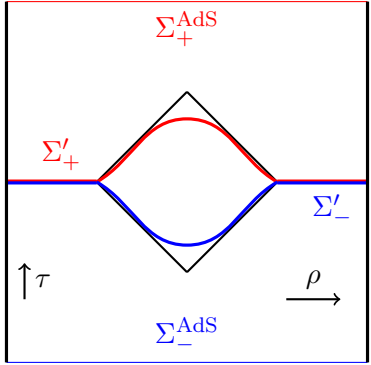

Having laid out the statement of AdS/CFT, we now explain how to perform a flat limit. At the level of the geometry, we need to take the AdS radius large such that the line element \eqref{eq:ds2} becomes that of Minkowski space. The explicit limit reads $\tau=t/L$ and $\rho=r/L$ so that
\be
ds^2 \xrightarrow[L \rightarrow\infty]{} -dt^2+dr^2+r^2d\Omega_{d-1}^2\, .
\ee
The operation can be regarded as zooming into the center region of AdS, where the geometry is locally flat. One can now think of AdS Cauchy slices $\Sigma'_{\pm}$ that match the slices $\Sigma_{\pm}$ in Minkowski space, but extend in AdS space as shown in figure \ref{fig:Limit}. In order to compute the S-matrix amplitude in Minkowski space, we need to choose the states at the slices $\Sigma_{\pm}$. In AdS, this can be achieved by exciting normalizable modes at $\Sigma_{\pm}^{\prime}$. 

We thus conclude that the S-matrix element in flat space is computed by a large $L$  limit of a CFT expectation value like the one appearing in equation \eqref{eq:NormPhiZ}. The last ingredient we need is a way to construct the CFT states $\vert {\text{in}} \rangle$ and  $\langle {\text{out}} \vert$  in terms of the field associated to a scattering state in $\Sigma_{\pm}$. For this, we will look for an operator $\hat{{\phi}}_{\text{in}}$ such that
\be
\hat{{\phi}}_{\text{in}} \vert 0 \rangle= \vert {\text{in}} \rangle\, .
\ee
We then need to recast the operator $\hat{{\phi}}_{\text{in}}$ defined on the Cauchy slice $\Sigma_-$ deep in the bulk in terms of a CFT operator at the boundary. Once this is done, the resulting S-matrix element will simply read
\be
\begin{split}
\langle \text{out}   \vert   \text{in}  \rangle =&\lim_{L\rightarrow\infty} \langle 0 \vert\hat{  {\phi}    }_{\text{out}}  e^{i \int d^d x \, \phi_0(x){\cal O}(x)} \hat{ {\phi}}_{\text{in}}\vert 0\rangle_{\phi_0=0}\, .
\end{split}
\ee
In the following section we will explicitly construct $\hat{{\phi}}_{\text{in/out}}$ in the CFT.

\subsection{Global reconstruction of scalar operators in AdS/CFT}
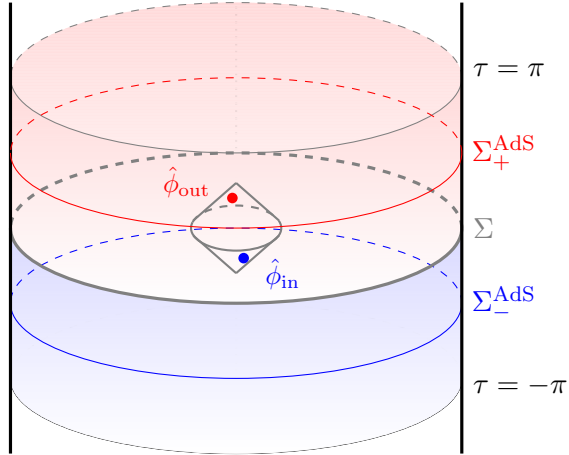
\begin{figure}[]
\centering
\begin{tikzpicture}
\draw[white,very thick] (5,2) arc (0:-180:3 and 3);
\draw[white,very thick,name path=TOP](1.25,7) arc (180:360:0.75 and 0.25);
\draw[gray, thick,dotted] (2,2) -- (2,7) node [pos=1,above=2]{{$ $}} node [pos=0,below=2]{{$ $}};

\draw[black] (-1,2) arc (-180:0:3 and 1);
\draw[black,dashed](5,2) arc (0:180:3 and 1);
\shade[top color=blue!15!white,opacity=0.85] (5,4) arc (0:180:3 and 1)   --  (-1,2) arc (-180:0:3 and 1)   -- cycle;
\shade[top color=blue!15!white,opacity=0.85]  (-1,2) arc (180:360:3 and 1)   --   (5,4) arc (0:-180:3 and 1)    -- cycle;
\shade[top color=red!15!white,opacity=0.85] (5,6) arc (0:180:3 and 1)   --  (-1,4) arc (-180:0:3 and 1)   -- cycle;
\shade[top color=red!15!white,opacity=0.85]  (-1,4) arc (180:360:3 and 1)   --   (5,6) arc (0:-180:3 and 1)    -- cycle;
\draw[gray] (-1,6) arc (-180:0:3 and 1);
\draw[gray,dashed](5,6) arc (0:180:3 and 1);

\draw[blue,dashed](5,3) arc (0:180:3 and 1) node[pos=0,right]{$\Sigma^{\text{AdS}}_-$};
\draw[blue](-1,3) arc (180:360:3 and 1);
\draw[gray,very thick,dashed](5,4) arc (0:180:3 and 1) node[pos=0,right]{$\Sigma$};
\draw[gray,very thick](-1,4) arc (180:360:3 and 1);
\draw[gray,thick](2.6-0.03,4+0.1)--(2,4.6);
\draw[gray,thick](1.4+0.03,4+0.1)--(2,4.6);
\draw[gray,thick,dashed](2.6,4) arc (0:180:0.6 and 0.3);
\draw[gray,thick](2.6,4) arc (0:-195:0.6 and 0.3);

\draw[gray,thick](2.6-0.03,4-0.1)--(2,3.4);
\draw[gray,thick](1.4+0.03,4-0.1)--(2,3.4);

\draw[black,very thick,-dotRed-=1] (2-0.05,4.35)--(2-0.05,4.4) node[pos=1,above=6,left=4]{\color{red}\small $\hat{{\phi}}_{\text{out}}$};
\draw[black,very thick,-dotBlue-=1] (2+0.1,3.65)--(2+0.1,3.60) node[pos=1,below=6,right=4]{\color{blue}\small $\hat{{\phi}}_{\text{in}}$};
\draw[black,very thick] (-1,1) -- (-1,7);
\draw[black,very thick] (5,1) -- (5,7) node [pos=0.85,right]{$\tau=\pi$} node [pos=0.15,right]{$\tau=-\pi$};
\draw[red,dashed](5,5) arc (0:180:3 and 1) node[pos=0,right]{$\Sigma^{\text{AdS}}_+$};
\draw[red](-1,5) arc (180:360:3 and 1);
\end{tikzpicture}
    \caption{The bulk field $\hat{\phi}(x)$ is placed inside a scattering region around the center of global AdS. The local operator can be reconstructed semi-classically in the boundary using HKLL. For the incoming field, the reconstruction involves  the shaded blue part of the boundary spanning $\tau\in(-\pi,0)$. For the outgoing field, we choose the shaded red region spanning $\tau\in(0,\pi)$.  
}\label{fig:HKLL}
\end{figure}

In the previous section we have explained how the S-matrix amplitude of a physical process in flat space-time can be computed by taking a flat limit of a CFT correlator. The operators in the CFT correlators are boundary representations of bulk operators living in late/early time slices of the Minkowski geometry, which is placed deep inside the bulk geometry. We will place the scattering region at $\tau=0$ at the center of global AdS. The bulk operator has been argued to be free in virtue of the assumption of asymptotic decoupling. This means that before and after the scattering region, a scalar operator must obey the Klein-Gordon equation on AdS,
\be
\left(  \nabla^2-m^2      \right)\hat{{\phi}} = 0\, , \quad \text{with} \quad m^2 L^2=\Delta(\Delta-d)\, .
\ee
The operator has also been argued to be associated to normalizable modes of bulk fields, and so it has the following fall-off behavior
\be
\hat{{\phi}}(\rho,x)\xrightarrow[\rho \to \frac \pi 2]{} (\cos \rho)^{\Delta}{\cal O}(x)\, .
\ee
The problem of reconstructing such an operator in the CFT has been analyzed extensively in the literature. In this paper we will use the HKLL reconstruction method \cite{Hamilton:2006az}. A nice feature of the HKLL method is that it allows to reconstruct bulk operators as operators in a region of the CFT covering the whole boundary in any global time interval of length $\Delta \tau = \pi$.  We will choose our scattering region to be around around the center of AdS at $\tau = 0$, such that all excitations are far separated unless they are in a small interval of size $\epsilon\sim{\cal O}(L)^{-1}$ around $\tau=0$. Since the time-like geodesics emanating from the scattering region will come close again after time $\pi$, the region of validity for the free field expansion is limited to $\tau = (\epsilon, \pi - \epsilon)$ for out-fields and $\tau = (-\pi + \epsilon, - \epsilon)$ for in-fields. For a picture of this set-up, see  figure \ref{fig:HKLL}. 

The details of the HKLL reconstruction method applied to scalars can be found in \cite{Hamilton:2006az} and have been summarized in appendix \ref{app:ScalarHKLL}. The structure of the result in AdS$_4$/CFT$_3$ reads
\be\label{eq:ScalarHKLL}
\hat{{\phi}}_{\text{AdS,0}}(\rho,x)={1\over \pi}\int_{{\cal  T}} d\tau'\, \int d^2 \Omega\, \left[   K_{+}(\rho,x;x') {\cal O}^+(x')+ K_{-}(\rho,x;x') {\cal O}^-(x')        \right]\, ,
\ee
where $K_{\pm}$ are kernels, defined such that $\hat{{\phi}}_{\text{AdS,0}}(\rho,x)$ obeys the Klein-Gordon equation and asymptotes to its dual operator $\mathcal O$ at the boundary, with positive/negative frequency modes ${\cal O}^{\pm}$. The regions of integration are ${\cal T}=(-\pi,0)$ for incoming and ${\cal  T}= (0, \pi)$ for outgoing states (see figure \ref{fig:HKLL}). The explicit form of the kernels reads 
\be
\begin{split}
K_{\pm}(\rho,x;x')=&\frac 1 {\mathcal N_{K,\Delta}}  \sum_{\kappa\in \mathbb{Z}}\sum_{l=0}^{\infty}\sum_{m=-l}^{m=l} e^{\pm i \omega_{\kappa,l}(\tau-\tau')} Y_l^m(\Omega)Y_l^m(\Omega')^* \sin^l\rho \cos^{\Delta}\rho \,\\
&\times {}_2F_1\left(  -\kappa,l+\Delta+\kappa;\Delta-{1\over 2} \Big\vert \cos^2\rho  \right)\, ,
\label{eq:kernels_sec_2}
\end{split}
\ee
where $\omega_{\kappa,l}=\Delta+l+2\kappa$ are the normalizable modes of a scalar field in AdS, and the normalization is given in \eqref{eq:ScalarNorma}. Note that in equation \eqref{eq:kernels_sec_2} the sum over modes $\kappa$ which usually runs only over positive integers has been extended to run over all integers. This is possible since the additional terms do not contribute to the integral \eqref{eq:ScalarHKLL}. In the next subsection, we will take a flat limit of equation \eqref{eq:ScalarHKLL} and extract creation/annihilation operators in Minkowski space.

The limit is different depending on whether the field is massive or massless. The reason for this is that the mass and the conformal dimension of the dual operator are related by 
\be
m^2 L^2=\Delta(\Delta-d)\, .
\ee
In the  large L limit, a massive field in flat space will arise if $\Delta$ also scales with $L$ as $\Delta=m L$. If otherwise $\Delta\sim {\cal O}(1)$, then in the flat limit the mass will vanish as $m^2\rightarrow \Delta(d-\Delta)/L^2\rightarrow 0$. For simplicity, we will simply choose $\Delta=d$ for the case of massless fields, such that the bulk operator is not just massless in the flat limit, but is exactly massless in anti-de Sitter space.

\subsubsection{Flat limit of HKLL operator: Massless}\label{sec:MasslessRec}
We start with the expression for a scalar bulk operator in AdS$_d$, formula \eqref{eq:ScalarHKLL} with $\Delta=d$. A scalar bulk operator in Minkowski space can then be obtained by taking the flat limit. This is done by reparametrizing the bulk coordinates $(\rho, \tau,\Omega)$ by 
\be\label{eq:MinkRegion}
\rho={r\over L}\, , \quad \text{and}\quad \tau={t\over L}\, ,
\ee
before sending $L$ to infinity.
Inside the Minkowski region, at very early/late times, we expect the field to be free due to asymptotic decoupling. This means that the field operator must have a representation as 
\be
\hat{{\phi}}(x)= \int {{d^3 \vec{p}}\over{(2\pi)^3}}{1\over\sqrt{2\omega_{\vec{p}}}}\left(     \hat{a}_{\vec{p}}e^{i p \cdot x}+    \hat{a}^{\dagger}_{\vec{p}}e^{-i p \cdot x}  \right)\, , \quad \text{with}\quad x\in \text{Mink}_{3+1}\, .
\ee
The creation/annihilation operators appearing here obey canonical commutation relations
\be
[a_{\vec{p}},a^{\dagger}_{\vec{k}}]= (2\pi)^3 \delta^{(3)}(\vec{p}-\vec{k})\, .
\ee
Since the decomposition of $\hat{{\phi}}(x)$ can be done at early and late times, there exist two sets of creation and annihilation operators, one for early and one for late times. They span the in- and out-scattering Fock spaces. Scattering states can be prepared as follows,
\be
\vert \vec{k},\text{in}\rangle = \sqrt{2\omega_{\vec{k}}}\, \hat{a}^{\dagger}_{\text{in},\vec{k}}\vert 0 \rangle \, , \quad \text{and}\quad \langle \vec{p}, \text{out}\vert=\langle 0 \vert  \sqrt{2\omega_{\vec{p}}}\, \hat{a}_{\text{out},\vec{p}}\, .
\ee
The creation/annihilation operators can be extracted from the position space field operator via
\be\label{eq:aFromPhi}
\begin{split}
\hat{a}_{\text{in/out},\vec{p}} =&\lim_{t\rightarrow \mp \infty}{i\over \sqrt{2\omega_{\vec{p}}}} \int_{\Sigma} d^3 \vec{x} \, e^{-i p\cdot x} \overleftrightarrow{\partial_0} \hat{{\phi}}(x)\, , \\
\hat{a}^{\dagger}_{\text{in/out},\vec{p}} =&\lim_{t\rightarrow\mp\infty}{-i\over \sqrt{2\omega_{\vec{p}}}} \int_{\Sigma} d^3 \vec{x} \, e^{i p\cdot x} \overleftrightarrow{\partial_0} \hat{{\phi}}(x)\, .
\end{split}
\ee
where $\overleftrightarrow{\partial}_0=\overrightarrow{\partial}_0-\overleftarrow{\partial}_0$. In order to represent these operators in the conformal field theory, we simply have to apply formulas \eqref{eq:aFromPhi} to the HKLL operator \eqref{eq:ScalarHKLL} placed in the Minkowski scattering region \eqref{eq:MinkRegion}.

The first step is to write the HKLL operator in the scattering region. The replacement $\tau=t/L$ in the large $L$ limit yields stationary solutions unless $\omega_{\kappa,l}$ scales with $L$. This means that fluctuating modes in Minkowski space arise from AdS modes with large values of $\kappa$. We will thus define 
\be
\Delta+l+2\kappa\equiv w L\, .
\ee
In the large $L$ limit, the sum over $\kappa$ can be traded for an integral over $w$ and the discrete spectrum of scalar field fluctuations in AdS becomes a continuum in the flat limit. The HKLL kernels can now be written explicitly in the Minkowski scattering region
\be 
\begin{split}
K_{\text{in},\pm}(r,t,\Omega;x')&=\frac{i \sqrt 3}{\pi^2 L} \int {dw\over w}\, e^{\pm i w t}  e^{i w L \left(  \mp(\tau' + {\pi \over 2})  \right)}\,\sum_{l,m} Y_l^m(\Omega)Y^m_l (\Omega')^*\, (\mp i)^{-l}j_l(w r)\, ,\\
K_{\text{out},\pm}(r,t,\Omega;x')&=\frac{(-i) \sqrt 3}{\pi^2 L} \int {dw\over w}\, e^{\pm i w t}  e^{i w L \left(  \mp(\tau' - {\pi \over 2})  \right)}\,\sum_{l,m} Y_l^m(\Omega)Y^m_l (\Omega')^*\, (\pm i)^{-l}j_l(w r)\, . 
\end{split}
\ee
The Bessel functions arise from the flat limit of the hypergeometric functions appearing in the AdS kernels. Applying now \eqref{eq:aFromPhi} together with
\be
e^{i \vec{p} \cdot \vec{x}}= 4\pi \sum_{l',m'}i^{l'} j_{l'}(\omega_{\vec{p}}r) Y^{m'}_{l'}(\hat{p})  Y^{m'}_{l'}(\Omega)^* \, ,
\ee
the final result reads
\be\label{eq:aMassless}
\begin{split}
\sqrt{2\omega_{\vec{p}}} \, a_{\text{in},\vec{p}}=& c_- \int^{0}_{-\pi } d\tau \,   e^{i \omega_p L \left( \tau+{\pi \over 2}\right)}  {\cal O}^- \left(  \tau , -\hat{p}  \right)\, , \\
\sqrt{2\omega_{\vec{p}}} \, a^{\dagger}_{\text{in},\vec{p}}=& c_+  \int^{0}_{-\pi } d\tau  \,  e^{-i \omega_p L \left(\tau+{\pi \over 2}\right) } {\cal O}^+ \left( \tau , -\hat{p}  \right)\, , \\
\sqrt{2\omega_{\vec{p}}} \, a_{\text{out},\vec{p}}=& c_+ \int^{\pi }_{0} d\tau  \,   e^{i \omega_p L \left( \tau-{\pi\over 2}\right)}  {\cal O}^- \left(  \tau  , \hat{p}  \right)\, , \\
\sqrt{2\omega_{\vec{p}}} \, a^{\dagger}_{\text{out},\vec{p}}=& c_- \int^{\pi }_{0} d\tau  \,  e^{-i \omega_p L  \left( \tau-{\pi\over 2}\right)} {\cal O}^+ \left( \tau, \hat{p}  \right)\, , \\
\end{split}
\ee
where $\hat{p}$ is the angle of the momentum $\vec{p}$, and $-\hat{p}$ stands for its anti-podal. We have also defined an over-all constant $c_\pm= \pm i \frac{\sqrt{24}}{L \omega_p^2}$. Note that the integrals over global time here are highly oscillatory as $L \to \infty$ unless operators are evaluated in windows of size ${\cal O}(1/L)$ at $\tau=\pm {\pi \over 2}$. This suggests that we can ignore contributions away from this region. Equation \eqref{eq:aMassless} is precisely the prescription presented previously in the literature \cite{Fitzpatrick:2011hu,Hijano:2019qmi}, which we have derived here from first principles.

We want to point out two important aspects of the above expressions. First, in the scattering states we are considering, the operator $\mathcal O$ has a free-field like expansion, c.f. appendix \ref{app:ScalarHKLL}. This implies that we can replace non-local operators $\mathcal O^\pm$ by local operators $\mathcal O$ in \eqref{eq:aMassless}, since the difference $\mathcal O - \mathcal O^{\pm}$ integrates to zero against the phase. In other words, the creation and annihilation operators can be constructed by smearing local operators. In the following we will nonetheless keep the positive/negative frequency superscripts. Whenever we talk about the location of $\mathcal O^\pm$ this should be understand as either replacing $\mathcal O^\pm$ by $\mathcal O$, or equivalently evaluating $\mathcal O^\pm(\tau, \hat p)$ in that location. 

Second, the equations in \eqref{eq:aMassless} naively look like they would vanish in the $L \to \infty$ limit. However, the large $L$ limit should only be taken after CFT correlators of positive/negative frequency operators have been calculated. These correlation functions contain pieces with diverge as the flat space limit is taken and the role of the integrals and additional factors in $L^{-1}$ in \eqref{eq:aMassless} is to extract particular divergences which correspond to S-matrix elements in flat space. This can be seen explicitly in some easy sample calculations, see appendix \ref{app:sample_calc}.


\subsubsection{Flat limit of HKLL operator: Massive}\label{sec:MassiveRec}
In the case of massive scalar operators, the calculation is very different due to the fact that the conformal dimension of the CFT operator now scales with $L$ such that
\be
m^2 L^2 =\Delta(\Delta-d)\, , \quad \Rightarrow \quad \Delta={d\over 2}+  m L +{\cal O}(L)^{-1}\, .
\ee
Like in the massless case, we will also need the AdS quantum number $\kappa$ to be large in order to obtain fluctuating modes in flat space. We thus have
\be
\Delta+l+2\kappa =w L\, .
\ee
The positive frequency kernels for in- and out-fields in the flat limit now read
\be
\begin{split}
K^{\text{out/in}}_{+}(r,t,\Omega;x')=& \frac{1}{2 \pi} \left(\frac{ML}{\pi^3}\right)^{\frac 1 4} L \int dw \,e^{i w t}\, k\, \left( {2m\over (\pm i) k}     \right)^{m L +{1\over 2}}\,    e^{i w L\left[  {\pi \over 2}+\frac i 2 \log\left(w+m\over w-m \right) \mp \tau'   \right]} \\
&\times\sum_{l,m} Y^{m}_{l}(\Omega) Y^{m}_{l}(\Omega')^*   \,{(\pm i)}^{-l}j_l(k r) \, ,
\end{split}
\ee
where we have defined 
\be
k\equiv \sqrt{w^2-m^2}\, .
\ee
The kernels for negative frequencies can be obtained by conjugation.
Plugging this expression in our formula of the HKLL operator and performing the Minkowski space coordinate integrals in \ref{eq:aFromPhi} yields for the out-state creation operator,
\be\label{eq:aMassive}
 \sqrt{2\omega_{\vec{p}}} \hat{a}^\dagger_{\text{out},\vec{p}} =  c_-(L,m,|\vec p |) \, \int_0^\pi d\tau' 
 e^{i \omega_{\vec{p}} L\left[  {\pi \over 2}+{i \over 2} \log\left(\omega_{\vec{p}}+m\over \omega_{\vec{p}}-m    \right) - \tau'   \right]}   {\cal O}^+(\tau',\hat{p})\, ,
\ee
where we have defined 
\begin{align}
c_\pm(L,m,|\vec p |) = \frac{1}{2 \pi} \left(\frac{mL}{\pi^3}\right)^{\frac 1 4}  \, \left( {2m\over (\mp i) \vert \vec{p}\vert }     \right)^{m L +{1\over 2}} L.
\end{align}
The annihilation operator can be obtained from this expression by complex conjugation. The expressions for the in-state operators are obtained analogous to the massless case by changing the integration region to $(-\pi,  0)$, changing the sign of $\frac \pi 2$ in the phase as well as $c_+ \leftrightarrow c_-$, and replacing $\hat p \to - \hat p$. 
Note that in the large $L$ limit, the real part of the exponential only vanishes, if
\be\label{eq:Imtau}
\text{Im}\left(   \tau'  \right)=\frac 1 2 \log\left( \omega_{\vec{p}}+m\over \omega_{\vec{p}}-m     \right)\, .
\ee
It is therefore convenient to move the $\tau$ contour such that it runs along this value in the complex plane.
Again, the integrand contains an exponential which highly oscillates as $L$ is sent to infinity. This focuses the integral around a window of size ${\cal O}(1/L)$ at Re$\left(\tau'\right)\sim {\pi \over 2}$. We conclude that creation operators in Minkowski space can be understood as large $L$ limits of operators with large conformal dimension placed in a conformal field theory at a complex value of global time\footnote{This was shown less rigorously in \cite{Hijano:2019qmi} by one of the authors. }. In the language of previous sections, such values of global time parametrize the Euclidean half spheres $\partial {\cal M}_{\pm}$ drawn in figure \ref{fig:AdS}.

\subsection{Mapping of asymptotic regions}\label{sec:Interpretation}
Formulas \eqref{eq:aMassless} and \eqref{eq:aMassive} allow for an identification of regions of the boundary CFT with the different asymptotic regions of Minkowski space-time. The oscillatory nature of the integrand in formula \eqref{eq:aMassless} suggests that light operators placed at small windows of global time around $\pm {\pi \over 2}$ correspond to massless particles in flat space. This region of the CFT plays the role of null infinity ${\cal I}^{\pm}$. Throughout the text, we will thus refer to these CFT regions as $\tilde{\cal I}^{\pm}$. The metric of the CFT in one of these fringes reads
\be
\label{eq:celestial_sphere_metric}
\tau= {\pi \over 2}+{u\over L} \quad\Rightarrow\quad ds^2_{{\cal I}^{\pm}}=L^2\left(  -d\tau^2+d\Omega^2  \right) = -du^2+L^2d\Omega^2\, .
\ee
In the large $L$ limit, this metric resembles the metric of asymptotic null infinity where $u$ plays the role of retarded time.

\begin{figure}[]
\centering
\begin{tabular}{cc}
\begin{subfigure}[t]{0.41\textwidth}
\centering
\begin{tikzpicture}[scale=0.8]
\shade[top color=white, bottom color=black!45!white,opacity=0.75]  (5,2) arc (0:180:3 and 1)  --  (5,2) arc (0:-180:3 and 3)   -- cycle;
\shade[top color=white, bottom color=gray,opacity=0.75]  (-1,2) arc (180:360:3 and 1)   --  (5,2) arc (0:-180:3 and 3)   -- cycle;
\draw[black,very thick] (5,2) arc (0:-180:3 and 3) node[pos=0.5,below]{$\partial{\cal M}_-$};

\draw[gray, thick,dotted] (2,2) -- (2,7) node [pos=1,above=2]{{$ $}} node [pos=0,below=2]{{$ $}};
\draw[blue,very thick,dashed,opacity=0,name path=TOPBD2](5,1.7) arc (0:180:3 and 1);
\draw[blue,very thick,dashed,name path=LOWBD](5,2) arc (0:180:3 and 1);
\draw[blue,very thick,dashed,opacity=0,name path=TOPBD](5,2.3) arc (0:180:3 and 1);
\tikzfillbetween[of=TOPBD and TOPBD2]{blue, opacity=0.5};

\draw[blue,very thick,opacity=0, name path=TOPB2](-1,1.7) arc (180:360:3 and 1);
\draw[blue,very thick, name path=LOWB](-1,2) arc (180:360:3 and 1);
\draw[blue,very thick,opacity=0, name path=TOPB](-1,2.3) arc (180:360:3 and 1);
\tikzfillbetween[of=TOPB and TOPB2]{blue, opacity=0.5};

\draw[black,very thick] (-1,2) -- (-1,7) node [pos=0,left]{\color{blue} $\tilde{\cal I}^-$} node [pos=1,left]{\color{red} $\tilde{\cal I}^+$};
\draw[black,very thick] (5,2) -- (5,7) node [pos=1,right]{$\tau={{\pi}\over 2}$} node [pos=0,right]{$\tau=-{{\pi}\over 2}$};

\draw[red,very thick,dashed,opacity=0,name path=TOPRD2](5,7.3) arc (0:180:3 and 1);
\draw[red,very thick,dashed,name path=TOPRD](5,7) arc (0:180:3 and 1);
\draw[red,very thick,dashed,opacity=0,name path=LOWRD](5,6.7) arc (0:180:3 and 1);
\tikzfillbetween[of=TOPRD2 and LOWRD]{red, opacity=0.5};

\draw[black,very thick] (5,7) arc (0:180:3 and 3) node[pos=0.5,above]{$\partial{\cal M}_+$};
\shade[top color=gray,bottom color=white,opacity=0.75]  (-1,7) arc (180:360:3 and 1)  --  (5,7) arc (0:180:3 and 3)  -- cycle;

\draw[red,very thick,opacity=0,name path=TOPR2](-1,7.3) arc (180:360:3 and 1);
\draw[red,very thick,name path=TOPR](-1,7) arc (180:360:3 and 1);
\draw[red,very thick,opacity=0,name path=LOWR](-1,6.7) arc (180:360:3 and 1);
\tikzfillbetween[of=TOPR2 and LOWR]{red, opacity=0.5};

\draw[cyan, very thick] (3.5,5.82) -- (3.5,5.82+0.6);
\draw[cyan, very thick, snake it,->] (2,4) -- (3,5.5);

\end{tikzpicture}
\caption{}
\end{subfigure}
& \, \, \, 
\begin{subfigure}[t]{0.41\textwidth}
\centering
\begin{tikzpicture}[scale=3]

\draw[gray, thick,dotted] (0,-1) -- (0,1);

\draw [white, thick] (-1,-1)--(1,1);
\draw [white, thick] (-1,1)--(1,-1);

\draw [blue, thick] (-1,0)--(0,-1) node[pos=1,below]{\color{black} $i^-$};

\draw [red, thick] (1,0)--(0,1) node[pos=0.5,right=3]{${\cal I}^+$};
\draw [red, thick,-dot-=0,-dotW-=0,-dotB-=1] (-1,0)--(0,1) node[pos=1,above]{\color{black} $i^+$};

\draw [blue, thick,-dot-=0,-dotW-=0,-dot-=1] (1,0)--(0,-1) node[pos=0.5,right=3]{${\cal I}^-$} node[pos=0,right] {\color{black} \, $i^0$};

\draw[cyan, very thick, snake it,->] (0,0) -- (0.42,0.42);


\end{tikzpicture}
\caption{}
\end{subfigure}
\end{tabular}
    \caption{ The red fringe of global time can be parametrized by $\tau= {\pi \over 2}+{u\over L}$, where $u$ plays the role of retarded time at future null infinity ${\cal I}^+$. Similarly, the blue fringe parametrizes past null infinity ${\cal I}^-$. In the main text, we refer to these regions as $\tilde{\cal I}^{\pm}$. The shaded caps are analytic continuations of the boundary CFT in the imaginary direction of global time, and they play the role of future/past infinity $i^{\pm}$. In the main text we refer to these caps as $\partial {\cal M}_{\pm}$. The rest of the CFT can be interpreted as space-like infinity. As an example, an outgoing massless particle in flat space can be constructed in AdS/CFT by smearing an operator over a fringe of global time at ${\cal I}^+$.
}\label{fig:Interpretation}
\end{figure}
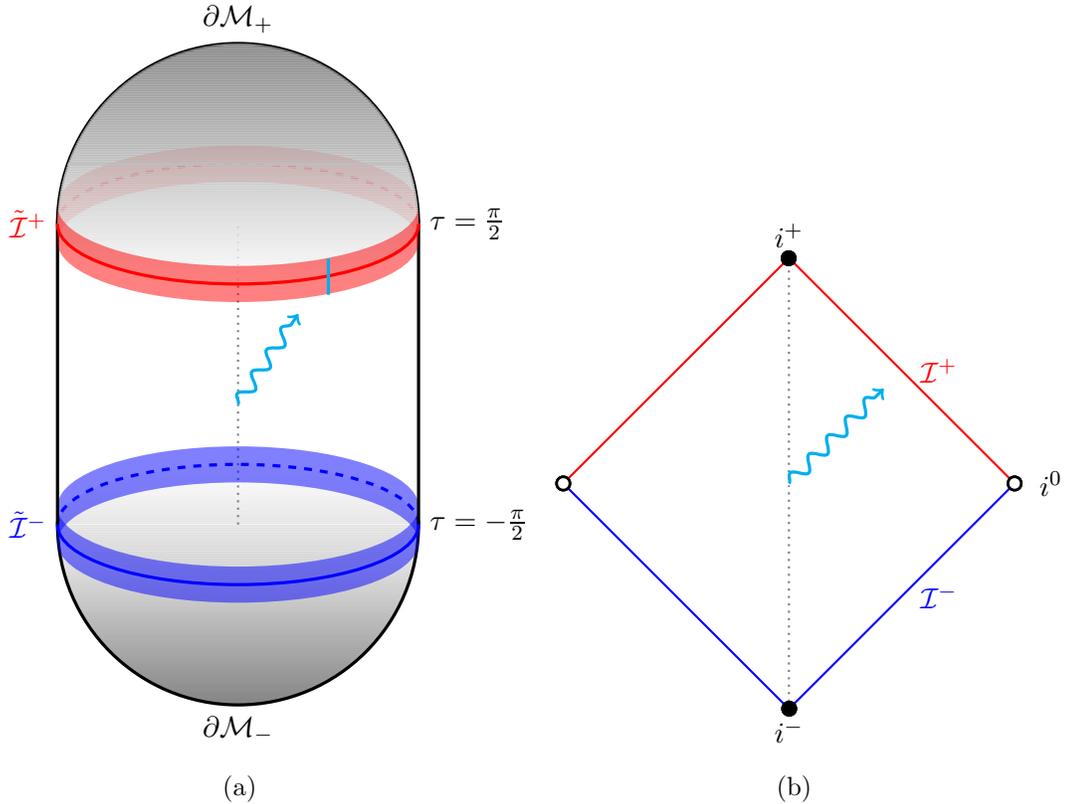

Equation \eqref{eq:aMassive} for massive scattering states also contains a highly oscillatory factor that suggests that heavy operators ($\Delta\sim L$) are located at complex values of global time $\tau=\pm{\pi \over 2}+i \tilde{\tau}$. This complexified region of the CFT plays the role of time-like infinity $i^{\pm}$. These regions have appeared in the text before (see figure \ref{fig:AdS}), and we have denoted them by $\partial{\cal M}_{\pm}$. At the level of the metric, we have
\be
ds^2_{\partial{\cal M}_{\pm}}=L^2\left(  -d\tau^2+d\Omega^2  \right) = L^2\left(  d\tilde{\tau}^2+d\Omega^2  \right) \, .
\ee
This metric can be written more suggestively by changing coordinates to
\be
R={1\over \sinh\tilde{\tau}}\, ,
\ee
such that the line element is now
\be
ds^2_{\partial{\cal M}_{\pm}}=L^2\left(    {dR^2\over R^2(1+R^2)} +d\Omega^2    \right)\, .
\ee
This metric is conformal to the metric of $H_3$ slices of Minkowski space-time used in \cite{deBoer:2003vf,Campiglia:2015qka,Kapec:2015ena}, and can be regarded as the metric of flat space at future/past infinity. Indeed, the special values of imaginary $\tau$ where operators must be placed in the CFT written in formula \eqref{eq:Imtau} correspond to precisely the values of $R$ that massive particles propagating in flat space pierce at $i^{\pm}$. Namely,
\be
R={\sqrt{\omega_{\vec{p}}^2-m^2}\over m}={\vert \vec{p}\vert\over m}\, .
\ee
The region in between $\tau=\pm {\pi\over 2}$ can be understood as space-like infinity $i^0$. For a picture of the structure analyzed in this section, see figure \ref{fig:Interpretation}.

Alternatively, we can interpret the  operators discussed in this work as operators in the celestial sphere. They take the form of Fourier transformations of local CFT operators. For example in the massless case we can formally define operators on the celestial sphere as
\be
\mathcal O_\omega(\hat p) \sim \lim_{L \to \infty} \frac 1 {L^2 \omega^2} \int_{-\infty}^{\infty} du \, e^{\pm i \omega u} \mathcal O \left(\frac u L \pm \frac \pi 2, \hat p \right).
\ee
This has some structural similarities with other proposals in which flat space holography is implemented by an infinite family of operators on the celestial sphere \cite{deBoer:2003vf,Lysov:2014csa,Cordova:2018ygx}.

\section{Photon scattering states}\label{sec:GaugeFields}
In this section we discuss how photon scattering states can be constructed in the flat limit of AdS/CFT. Asymptotic decoupling implies that photons obey the free Maxwell equation at early and late times. For the purpose of constructing scattering amplitudes, it is therefore sufficient to discuss the reconstruction of free gauge fields in AdS. This will be done explicitly in sections \ref{sec:FLM} and \ref{sec:FLE}. In section \ref{sec:Coulomb} we will discuss the back-reaction of charged particles on the gauge field, which will play a role when discussing Ward identities. 

Like in the scalar case, we need to understand how bulk path integrals in the presence of gauge fields compute generating functionals of CFT correlators. In the case of gauge fields, the problem becomes more subtle. In general, the solution to the field equations consists of two types of solutions which are characterized by their fall-off behavior at the asymptotic boundary. In the case of gauge fields both types of solutions fall off fast enough to be normalizable. The duality thus admits different formulations depending on what conditions we \emph{choose} to impose at the boundary. This leads to a rich and interesting amount of physics, some of which we will discuss throughout this section.

\subsection{Inequivalent quantization schemes of scalar fields in AdS}\label{sec:ScalarLagrange}
Before considering gauge fields, we start by briefly reviewing the analogous scenario involving scalar fields.
The issue of having two types of solutions which are both normalizable is also present in the case of scalar fields for a particular range of bulk masses \cite{Breitenlohner:1982jf}. In the range
\be\label{eq:MassRange}
1-{d^2\over 4}>m^2 L^2>-{d^2\over 4}\, ,
\ee
there are two inequivalent ways to quantize a scalar field in AdS. The general solution for a scalar field close to the boundary of AdS reads
\be
\phi(\rho,x) \xrightarrow[\rho \to {\pi \over 2}]{} (\cos\rho)^{\Delta_+}\alpha(x)+(\cos\rho)^{\Delta_-}\beta(x)\, .
\ee
Outside the mass range \eqref{eq:MassRange}, the mode associated to $\beta(x)$ is non-normalizable due to its behavior close to the AdS boundary. This leads to the ``standard'' construction of section \ref{sec:ScatteringStatesScalar}, where $\beta(x)$ becomes a non-dynamical source at the boundary that we called $\phi_0(x)$. Otherwise, when the bulk mass is in the range \eqref{eq:MassRange}, we could instead choose to fix $\alpha(x)$ at the boundary, and quantize the part of the field that asymptotes to $\beta(x)$. This leads to a holographic correspondence with a different conformal field theory \cite{Klebanov:1999tb}. More precisely, if we fix $\alpha(x)$ at the boundary, the CFT contains an operator of conformal dimension $\Delta_+$. If instead we fix $\beta(x)$, then the CFT contains an operator with conformal dimension $\Delta_-$. These two theories are related by renormalization group flow \cite{Witten:2001ua,Minces:2002wp}. The theory with an operator ${\cal O}_{\Delta_-}$ can be deformed by the relevant operator ${\cal O}_{\Delta_-}^2$, leading to a mixed boundary condition of the form $\alpha=f \beta$.

An interesting fact is that the two different conformal field theories are related by a Legendre transformation \cite{Klebanov:1999tb}. In this case, the CFT with non-dynamical source $\alpha(x)=J(x)$ is modified by making $J(x)$ dynamical and coupling it to a new background source $J'(x)$. The resulting CFT path integral corresponds to the bulk theory with the alternative boundary conditions $\beta(x)=J'(x)$. 

The take-away message from this discussion is that different choices of boundary conditions for our bulk fields lead to different conformal field theories that are related by a Legendre transformation. This is also the case for gauge fields, in which the relation between the two different boundary theories can be understood in the bulk as the statement of electric-magnetic duality.

\subsection{Electric-magnetic duality and $SL(2,\mathbb{Z})$ action on CFT$_3$'s}\label{sec:SL2Z}
Let us now turn to the case of a bulk theory with $U(1)$ gauge symmetry. As we will now review, there are an infinite number of such theories mapping onto each other under $SL(2,\mathbb{Z})$ transformations, which are an extension of electric-magnetic duality \cite{Witten:1995gf}. In terms of the dual description, the $SL(2,\mathbb Z)$ relates different conformal field theories, similar to the scalar case discussed above.

Consider a general solution for a gauge field in AdS$_4$ with boundary behavior
\be\label{eq:Aasymptotics}
{\cal A}_{\mu}(\rho,x) \xrightarrow[\rho \to \frac \pi 2]{}(\cos\rho)^1 \alpha_{\mu}(x)+(\cos\rho)^0 \beta_{\mu}(x)\, .
\ee
We could choose to fix $\beta_{\mu}(x)=A_{\mu}(x)$ at the boundary, which introduces a non-dynamical source $A_{\mu}(x)$ in the CFT. In this case $\alpha_{\mu}(x)$ corresponds to the finite energy bulk excitations spanning the low energy Hilbert space. Bulk path integrals now correspond to a generating functional of CFT correlators where the source $A_{\mu}(x)$ couples to a conserved current operator $j^{\mu}(x)$ with conformal dimension $\Delta_j=d-1=2$. Note that this boundary condition fixes the magnetic component of the gauge field at the boundary $M_i={1\over 2}\epsilon_{abc}F^{bc}$. We will thus refer to these boundary conditions as ``magnetic''. 

Another possibility is to fix the value of the electric field at the boundary $E_a=F_{\rho a }$ in order to obtain ``electric'' boundary conditions. This corresponds to fixing a boundary value of $\alpha_{\mu}(x)=B_{\mu}(x)$, while $\beta_{\mu}(x)$ is allowed to fluctuate. In this case, the conformal theory couples the source $B_{\mu}(x)$ to a conserved current constructed out of an operator with conformal dimension $\Delta_A=1$, which can be understood as a dynamical gauge field in the conformal theory.

The inequivalent CFTs described here are related through RG flow like in the scalar case \cite{Marolf:2006nd}. There is also a relation involving the $S$-transformation \cite{Witten:1995gf} of $SL(2,\mathbb Z)$. This is similar to the Lagrange transform that relates the two conformal theories dual to different quantization schemes for a bulk scalar field (see section \ref{sec:ScalarLagrange} above). The $S$-transformation acts as follows. We start with a CFT with Lagrangian ${\cal L}(\phi)$ and global $U(1)$ symmetry. Here, $\phi$ stands for a field in the CFT. Such a Lagrangian is sourced by the magnetic boundary condition,
\be
{\cal L}'(\phi)={\cal L}(\phi) + A_{\mu} j^{\mu}\, ,
\ee
where $j^{\mu}$ is the current operator associated to the global U(1) symmetry. CFT path integrals involving this Lagrangian compute bulk path integrals with magnetic boundary conditions,
\be
\langle e^{i \int_{\partial \text{AdS}} d^3 x\, A_{\mu}j^{\mu}} \rangle_{\text{CFT}} = \int [{\cal D}{\cal A}]_{A_{\mu}}\, e^{i S[{\cal A}]}\, .
\ee
where for simplicity we have omitted the specification of in/out CFT states and the corresponding conditions for bulk fields at space-like Cauchy slices. 
 The $S$-transformation of the conformal theory consists of regarding the background connection $A_{\mu}$ as dynamical. A current $\tilde{j}^{\mu}=\epsilon^{\mu\nu\rho}\partial_{\nu}A_{\rho}$ can then be introduced and coupled to a new background field $B_{\mu}$, such that the CFT Lagrangian reads
\be
{\cal L}(\phi,A)={\cal L}(\phi) + A_{\mu} j^{\mu}+B_{\mu}\tilde{j}^{\mu}\, .
\ee
Path integrals involving this Lagrangian will now compute bulk path integrals involving the electric boundary conditions $B_\mu$ described above, instead of the magnetic ones \cite{Yee:2004ju}. Explicitly, we can write
\be
\langle e^{i \int_{\partial \text{AdS}} d^3 x\, B_{\mu}\tilde{j}^{\mu}} \rangle_{\text{CFT}} = \int [{\cal D}{\cal V}]_{B_{\mu}}\, e^{i S'[{\cal V}]}\, ,
\ee
where the path integral in the CFT runs over $A_{\mu}$, and the bulk action $S'[{\cal V}]$ is the same as $S[{\cal A}]$ but with coupling $\tilde{e}=4\pi/e$. As shown in \cite{Yee:2004ju}, gauge fields with magnetic and electric boundary conditions are related through Hodge duality. The gauge connection with magnetic boundary conditions ${\cal A}$ is related to the gauge connection with electric ones ${\cal V}$ by
\be\label{eq:Hodge}
({\cal F}_{\cal A})_{\mu\nu} = {e^2\over 8 \pi} \epsilon_{\mu\nu\rho\sigma}({\cal F}_{\cal V})^{\rho \sigma}\, .
\ee
Alternatively, the above can be understood as fixing the boundary condition to be of magnetic type and making a choice as to whether the gauge potential of the field strength or that of the dual field strength are related to simple local operators in the boundary theory. 

The full $SL(2,\mathbb Z)$ is the natural extension of the above. In terms of boundary conditions, the $SL(2,\mathbb Z)$ theory relates magnetic boundary conditions to more general, Robin-type ones. In terms of the relation of bulk degrees of freedom, it acts as a map between gauge potentials which are related to simple operators in the boundary theory. In the following, we will only be interested in magnetic and electric boundary conditions, i.e., the electric-magnetic duality subgroup of $SL(2,\mathbb Z)$ generated by $S$.

\subsection{Flat limit: Magnetic boundary conditions}\label{sec:FLM}
In this section we will explore the standard choice of boundary conditions for the bulk gauge field, where we fix the magnetic field at the boundary to zero. Setting $\beta_{\mu}(x)=0$ means there is no magnetic field at the boundary, which forbids the existence of bulk magnetic charge. In a flat limit, we thus expect to recover a theory of electromagnetism without magnetic monopoles. We will follow the same logic as in the scalar case of section \ref{sec:ScatteringStatesScalar}. The boundary conditions on the holographic boundary determine the source, while boundary conditions at Cauchy slices $\Sigma_{\pm}$ specify a choice of CFT state. Instead of constructing scalar creation/annihilation operators, we now construct photon operators. 

Like in the scalar case, we will invoke the principle of asymptotic decoupling, which implies that our quantum fields are free away from the scattering region. In this section, we will study the radiative part of the gauge fields, and relegate coulombic contributions to section \ref{sec:Coulomb}. Free $U(1)$ gauge fields in Minkowski space can be written as follows
\be
\hat{\cal A}_{\mu}(x)=\int {d^3 \vec{q}\over (2\pi)^3} {1\over{\sqrt{2\omega_{\vec{q}}}}} \sum_{\lambda=\pm}\left(
\varepsilon^{(\lambda)}_{\mu} \hat{a}^{(\lambda)}_{\vec{q}} \, e^{i q\cdot x}
+
\varepsilon^{(\lambda)*}_{\mu} \hat{a}^{(\lambda)\dagger}_{\vec{q}} \, e^{-i q\cdot x}
\right)\, , \quad \text{with}\quad x\in \text{Mink}_{3+1}\, .
\ee 
In this expression $\lambda$ stands for the two different polarizations, and $\varepsilon^{(\lambda)}_{\mu}$ are null polarization vectors. We will be working in Lorenz gauge, so $\varepsilon_{\mu}q^{\mu}=0$, and we normalize these vectors such that $\varepsilon^{(+)}_{\mu}\varepsilon^{(-),\mu}=1$. Creation and annihilation operators can thus be extracted from the position space field using
\be\label{eq:aFromA}
\begin{split}
\hat{a}_{\vec{q}}^{(\lambda)} =&\lim_{t\rightarrow\pm\infty}{i\over \sqrt{2\omega_{\vec{q}}}} \int d^3 \vec{x} \,  (\varepsilon^{(\lambda),\mu})^*  e^{-i q\cdot x} \overleftrightarrow{\partial_0} \hat{{\cal A}}_{\mu}(x)\, , \\
\hat{a}^{(\lambda)\dagger}_{\vec{q}} =&\lim_{t\rightarrow\pm\infty}{-i\over \sqrt{2\omega_{\vec{q}}}} \int d^3 \vec{x} \, \varepsilon^{(\lambda),\mu} e^{i q\cdot x} \overleftrightarrow{\partial_0} \hat{{\cal A}}_{\mu}(x)\, ,
\end{split}
\ee
where $t \to \pm\infty$ for out- and in-states, respectively. In order to write these expressions in terms of CFT operators, we need an expression for the operator $\hat{\cal A}_{\mu}(x)$ in the asymptotic parts of the Minkowski scattering region located deep inside the bulk. In the scalar case we achieved this using HKLL reconstruction. The same can be done for gauge fields. In \cite{Heemskerk:2012np}, a representation of bulk gauge fields in the AdS Poincar\'e patch was derived. Here we are interested in results in global AdS. The reconstruction of a Maxwell field in global AdS is given in appendix \ref{app:MaxfieldReconstruction}, and to our knowledge has not been presented previously. We quote the main results here. As explained above, our gauge field with magnetic boundary conditions is dual to a current operator in the CFT. In this section however, we are only taking the radiative part of the field into account, in line with scattering theory. This means that the dual operator is not the full current, but only its radiative part, which we could denote by $j_{\mu}^{\text{R}}$. In order to avoid cluttering of notation, we will simply write $j_{\mu}$ in what follows. We thus look for a field operator with boundary behavior
\be
{\cal A}_{\mu}(\rho,x)\xrightarrow[\rho \to \frac \pi 2]{} \cos\rho \, j_{\mu}(x)\, ,
\ee
up to a normalization constant which depends on the normalization of the CFT current two point function.
It can be written in Lorenz gauge as
\be
{\cal A}_{\mu}(\rho,x)={1\over \pi}\int_{\mathcal T}d\tau'\, \int d^2\Omega' \, 
\left[
K^V_{\mu}(\rho,x;x') \epsilon_{\tau' a b}\partial^{a}j^{b}  (x')
+
K^S_{\mu}(\rho,x;x')  \partial_{a}j^{a}(x')
\right]+\text{h.c}
\, .
\ee
Here, the Latin indices run over the spatial coordinates of the $S^2$. The explicit form of the kernels $K^V$ and $K^S$ can be found in appendix \ref{app:MaxfieldReconstruction}. The reconstruction region $\mathcal T$ is $\tau \in (-\pi, 0)$ for incoming and $\tau \in (0,\pi)$ for outgoing particles. We now take the flat limit by placing the gauge field in the Minkowski scattering region
\be
\tau={t\over L}\, , \quad \text{and}\quad \rho={r\over L}\, , \quad \text{with}\quad L\rightarrow \infty\, .
\ee
Like in the scalar case, finite energy states in Minkowski space correspond to large AdS quantum numbers. The reconstructed operator in the large $L$ limit becomes
\be
\begin{split}
{\cal A}_{\mu}=& -{1\over 8 \pi} \int d^3 x'\, \int   dk\, \sum_{l,m}{ Y_l^m(\Omega')^* \over{-l(l+1)}}  e^{i k L \left({\pi \over 2}-\tau'\right)} e^{i k t}\\
& \left[   \epsilon_{\tau'}^{\,\, a b}\nabla_a j^+_b(x') \, \left(\vec{r}\times{d\over d\vec{r}}\right)+{i\over k} \nabla^a j^{+}_a(x') \vec{\nabla}\times\left( \vec{r}\times{d\over d\vec{r}}\right) \right] i^{-l} j_l(k r) Y_l^m(\hat{\Omega})\\
&+\text{h.c} \, .
\end{split}
\ee
This formula now constructs a position space gauge field in the Minkowski scattering region. This result can then be inserted into equation \eqref{eq:aFromA} to obtain expressions for photon creation/annihilation operators. After a bit of fun algebra, we obtain
\be
\begin{split}
\sqrt{2\omega_{\vec{q}}}\, \varepsilon^{(\lambda)}_{\mu} \, a_{\vec{q}}^{\dagger(\lambda)}=& -  {1\over 8\omega_q}   \int d^3x'\, e^{i k L  \left(  {\pi \over 2}-\tau'  \right)    }    \left[   \epsilon_{\tau'}^{\,\, a b}\nabla_a j^+_b(x') \, X_{\mu}- \nabla^{a}j^{+}_a(x') Y_{\mu} \right]  \, ,
\end{split}
\ee
where we have defined the four-vectors 
\be\label{eq:XY}
\begin{split}
X=\left(\vec{q}\times{d\over d\vec{q}}\right) G(\Omega',\hat{q})\, , \quad \text{and}\quad
Y=&\left[\left( \vec{q}\times{d\over d\vec{q}}\right)G(\Omega',\hat{q})\right]\times {\vec{q}\over \omega_q}   \, .
\end{split}
\ee
Here, the distribution $G(\Omega',\hat{q})$ is the Green's function associated to the Laplace operator on the $S^2$. The result can be written in much simpler form by integrating by parts along the sphere. We will be using a complex parametrization of the coordinates on $S^2$, as explained in appendix \ref{app:coordS2}. In these coordinates, the direction of the momentum of the photon is given by $(z_q,\bar{z}_q)$. For out state creation operators, the final result is
\be\label{eq:MagneticBCphotons}
\begin{split}
\sqrt{2\omega_{\vec{q}}} \,   a_{\vec{q}}^{\dagger(-)}=&   {-1\over 4\omega_q}{1+z_q \bar{z}_q\over{\sqrt{2}}}  \int d\tau' \,  e^{i \omega_{\vec{q}} L\left({\pi\over 2}-\tau' \right)}
\int d^2 z' {1\over (z_q-z')^2}j^+_{\bar{z}'}(\tau',z',\bar{z}')
\, , \\
\sqrt{2\omega_{\vec{q}}} \,  a_{\vec{q}}^{\dagger(+)}=&   {-1\over 4\omega_q}{1+z_q \bar{z}_q\over{\sqrt{2}}}  \int d\tau' \,  e^{i \omega_{\vec{q}} L\left({\pi\over 2}-\tau' \right)}
\int d^2 z' {1\over (\bar{z}_q-\bar{z}')^2}j^+_{z'}(\tau',z',\bar{z}')\, .
\end{split}
\ee
The annihilation operators, as well as the operators acting on the in-state Fock space can again be obtained by using the prescription given below \eqref{eq:aMassive}. The factor $c$ that appeared in equation  \eqref{eq:aMassive} is set to $1$, which fixes the normalization of the current in the CFT.
We conclude that asymptotic photon scattering states can be prepared in Minkowski space by smearing current operators over the sphere as in the formulas above. Like in the massless scalar case, the integrals appearing in \eqref{eq:MagneticBCphotons} are focused around a small window of global time at $\tau'={\pi \over 2}$. This again shows how this window acts like asymptotic null infinity of flat space that is pierced by photons in scattering experiments in flat space-time.

\subsection{Flat limit: Electric boundary conditions}\label{sec:FLE}
In the previous subsection we have argued that one can set up scattering events involving photons in flat space by considering the flat limit of the dual of a holographic conformal field theory and smearing conformal currents over the sphere. As explained in section \ref{sec:SL2Z}, we have made use of magnetic boundary conditions in AdS that forbid the existence of magnetic charges in the bulk. In this section we will consider boundary conditions that fix the electric field to vanish at the AdS boundary, such that electric charge is forbidden instead. 

We thus look for a new dynamical bulk gauge field ${\cal V}_{\mu}$ with asymptotic behavior given by
\be
{\cal V}_{\mu} \xrightarrow[\rho \rightarrow \frac \pi 2]{} A_{\mu}\, ,
\ee
again up to normalization.
Such a bulk field can be reconstructed at the boundary using a similar calculation to the one used in the previous section. The explicit calculation can be found in appendix \ref{MaxfieldReconstructionElectric}. Taking a flat limit as described in the previous section yields the following expression for a gauge field in Minkowski space in terms of CFT operators
\be
\begin{split}
{\cal V}_{\mu}=&{L \over 8} {1\over  \pi} \int d^3 x'\, \int dk\, \sum_{l,m}{ Y_l^m(\hat{\Omega}')^* \over{-l(l+1)}} e^{i k L\left({\pi \over 2}-\tau'\right)}  e^{i k t}  i k\\
&\left[  \, \epsilon_{\tau'}^{\,\, a b}\nabla_a A^+_b(x') \, \left(\vec{r}\times{d\over d\vec{r}}\right)-{i \over k}\nabla^{a}A^{+}_a(x') \vec{\nabla}\times\left( \vec{r}\times{d\over d\vec{r}}\right) \right]  i^{-l} j_l(k r) Y_l^m(\hat{\Omega}) \, \\
&+\text{h.c}\, .
\end{split}
\ee
From this expression we can extract creation operators using formulas \eqref{eq:aFromA}, this time applied to the connection ${\cal V}_{\mu}$. The result reads
\be
\begin{split}
\sqrt{2\omega_{\vec{q}}}\, \varepsilon^{(\lambda)}_{\mu} \, v_{\vec{q}}^{\dagger(\lambda)}=& -  i{L\over 8}   \int d^3x'\, e^{i k L  \left(  {\pi \over 2}-\tau'  \right)    }    \left[   \epsilon_{\tau'}^{\,\, a b}\nabla_a A^+_b(x') \, X_{\mu}+\nabla^{a}A^{+}_a(x') Y_{\mu} \right]  \, ,
\end{split}
\ee
where $X$ and $Y$ were defined in formula \eqref{eq:XY} above. Integrating by parts along the coordinates of the sphere yields two dimensional delta functions on the $S^2$, leading to a very simple result\footnote{The reconstruction written here has been performed in terms of time derivatives of the operators $A_{\mu}$ for future convenience, but expressions in terms of the operators alone can also be written.}
\be\label{eq:ElectricBCphotons}
\begin{split}
\sqrt{2\omega_{\vec{q}}} \,   v_{\vec{q}}^{\dagger(-)}=&{1\over 4\omega_q}{1+z_q \bar{z}_q\over{\sqrt{2}}}  \int d\tau' \,  e^{i \omega_{\vec{q}} L\left({\pi\over 2}-\tau' \right)}
\partial_{\tau'}A^+_{z'}(\tau',z_q',\bar{z}_q')
\, , \\
\sqrt{2\omega_{\vec{q}}} \,  v_{\vec{q}}^{\dagger(+)}=&{1\over 4\omega_q}{1+z_q \bar{z}_q\over{\sqrt{2}}}  \int d\tau' \,  e^{i \omega_{\vec{q}} L\left({\pi\over 2}-\tau' \right)}
\partial_{\tau'}A^+_{\bar{z}'}(\tau',z_q',\bar{z}_q')\, .
\end{split}
\ee
The construction of the other operators follows as in the magnetic case.
We conclude that ``dual'' photons in flat space can be prepared in the conformal field theory by simply inserting operators $A_{\mu}$ in a window of global time at $\tau'={\pi \over 2}$ at a single point on the sphere specified by the angle of the photon's momentum $\vec{q}$.

\subsection{Coulombic source terms}\label{sec:Coulomb}
In this section we have reconstructed bulk operators associated with radiative modes of a gauge field and the operators appearing in formulas \eqref{eq:MagneticBCphotons} and \eqref{eq:ElectricBCphotons} capture just those contributions. This has been achieved by finding specific kernels obeying the homogeneous Maxwell equations. The bulk theory can however generally contain charged matter. For example, the scalar particles considered in the previous sections could be complexified and charged under the $U(1)$ symmetry. This implies the existence of sources in the Maxwell equations, so that we would have
\be
\nabla_{\mu}\hat{F}^{\mu\nu}=\hat{J}^{\nu}\, ,
\ee
where the current operator $\hat{J}^{\mu}$ can be constructed explicitly out of the scalar bulk operators. In the context of scattering theory, interaction of the electromagnetic fields with sources is ignored in the asymptotic regions. This is typically justified by arguing for asymptotic decoupling. This also justifies our construction. In the next subsection, however, we will demonstrate the conservation of the charges of asymptotic gauge transformations. In the original flat space argument \cite{He:2014cra,Kapec:2015ena} -- as well as in our argument -- the derivation of those charges relies on the knowledge of the correct Li\'enard-Wiechert fields. Those contribute to non-radiative gauge field modes in the bulk.

For the purpose of reconstructing asymptotic gauge fields, in addition to the radiative reconstruction presented previously, our gauge field now also has a Coulombic (C) term like
\be\label{eq:ACoulomb}
\hat{A}^{(\text{C})}_{\mu}(X) = \int d^{d+1}Y \, G_{\mu,\nu}(X\vert Y) \hat{J}^{\nu}(Y)\, ,
\ee
where $G_{\mu,\nu}$ is a bulk-to-bulk propagator in Lorenz gauge. Note that the expression above can be written explicitly in the CFT by using the scalar bulk operator reconstruction formulas derived in appendix \ref{app:ScalarHKLL}. Because the  current operator $ \hat{J}^{\nu}$ is quadratic in the bulk scalar operator, formula \eqref{eq:ACoulomb} involves double trace operators constructed from the CFT operator dual to the bulk scalar field. 

At the boundary, the Coulombic gauge field \eqref{eq:ACoulomb} results in an additional contribution to the boundary current
\be
\lim_{\rho\rightarrow {\pi \over 2}} \hat{A}^{(\text{C})}_{\mu}(\rho, x)  = \cos\rho \, j_{\mu}^{(\text{C})}(x)\, .
\ee
Expectation values of the current operator $j_{\mu}^{(\text{C})}$ compute Li\'enard-Wiechert potentials at the boundary of Anti-de Sitter space-time. This part of the current is also responsible for non-trivial Ward-Identities in the conformal field theory, wich we will use in the next section to study soft theorems in Minkowski space. The conserved current operator dual to the asymptotic gauge field thus reads
\be
j_{\mu}(x)=j^{\text{R}}_{\mu}(x)+j^{\text{C}}_{\mu}(x)\, ,
\ee
where $j^{\text{R}}_{\mu}(x)$ is the radiative part of the current appearing in formula \eqref{eq:MagneticBCphotons}. A similar story can also be told for the dynamical gauge field playing a role in the reconstruction of bulk fields with electric boundary conditions.


\section{From conformal Ward identities to Weinberg soft theorems}\label{sec:WST}
In this note we have established a relation between operators in conformal theories and creation/annihilation operators in flat space-time. We can now ask what the implications of this connection are. The objective of this section is to show how Ward identities in a three dimensional conformal field theory are related to Weinberg soft theorems in Minkowski space-time in four dimensions. We start with a discussion on currents and Ward identities in conformal field theories with global $U(1)$ symmetry (and their S-transformations). 


\subsection{Ward identities in CFT}
A bulk AdS$_4$ theory containing a $U(1)$ gauge field with ``magnetic" boundary conditions is equivalent to a conformal field theory with global $U(1)$ symmetry. The charge under which the operators ${\cal O}$ in the CFT transform has an associated conserved current operator $j_{\mu}$, which is dual to the bulk gauge field. As explained in section \ref{sec:Coulomb}, such a current has two pieces; One comes from the radiative part of the gauge field, while the other one captures Coulombic contributions. Even though the total current is transverse, time ordered expectation values are only conserved up to contact terms\footnote{
This is especially easy to see in the derivation of the Ward identity in the operator formalism.
Time ordered correlators can be defined as
\be
\langle 0\vert T\{   j_{\mu}(x) {\cal O}(x_1)\cdots \}\vert 0 \rangle = \langle  0\vert  j_{\mu}(x) {\cal O}(x_1)\cdots \vert 0\rangle \Theta(t>t_1>\cdots)+\text{other orderings}.
\ee
The divergence of this correlator exhibits contact terms. When the divergence hits the current operator, the result is trivial as the current is conserved. However when the derivative hits the multi-argument step-function, terms involving the commutator $[j_0(x),{\cal O}(x_j)]$ appear. These are non-trivial due to the Coulombic part of the current $j^{(\text{C})}_{\mu}$ discussed in section \ref{sec:Coulomb}.
}, which is made explicit by the Ward identities. 
\be\label{eq:EQ1IBP}
\begin{split}
& \partial_{\mu} \langle 0 \vert     T\{   j^{\mu}(x)   {\cal O}(x_1)\cdots  {\cal O}(x_n)\bar{ {\cal O}}(y_1)\cdots  \bar{ {\cal O}}(y_m)        \}                 \vert  0 \rangle \\
&= \left(\sum_{i=1}^n q_i \delta^{(3)} (x-x_i)- \sum_{j=1}^m   q_j \delta^{(3)} (x-y_i)\right)
\langle 0 \vert    T\{     {\cal O}(x_1)\cdots  {\cal O}(x_n)\bar{ {\cal O}}(y_1)\cdots  \bar{ {\cal O}}(y_m)        \}                 \vert  0 \rangle
\, .
\end{split}
\ee
This formula will be used later to make a statement about the asymptotic symmetries of QED in flat space-time.

Consider now allowing for a non-vanishing magnetic field at the boundary. This way, magnetic monopoles can be introduced into the theories we are working with \cite{Sachdev:2012tj,Pufu:2013eda}. They consist of background connections $A_{\mu}$ that cannot be well-defined globally. From the background connection we can define a background field strength and its Hodge dual
\be
f_{\mu\nu}=\partial_{\mu}A_{\nu}-\partial_{\nu}A_{\mu} \, \quad \text{and}\quad (*f)^{\mu}={1\over 2}\epsilon^{\mu\nu\rho}f_{\nu\rho}\, .
\ee
Magnetic monopoles are then background connections violating the Bianchi identity
\be
\partial_{\mu}(*f)^{\mu}=\chi \, ,
\ee
where $\chi$ is a source (see \cite{MouraMelo:2000zc,Abreu:2001dq} for an analysis on the space of solutions). One can think of these monopoles as providing non-trivial charge to the topological current $(*f)^{\mu}$.

After $S$-transformation of the CFT discussed in section \ref{sec:SL2Z}, the field $A_{\mu}$ is dynamical, and $(*f)^{\mu}$ is the conserved current of the theory. The breaking of the Bianchi identity can then be written as a Ward identity associated to $(*f)^{\mu}$,
\be\label{eq:EQ2}
\begin{split}
& \partial_{\mu}\langle 0 \vert    T\{  (*f)^{\mu}(x)   {\cal M}(x_1)\cdots {\cal M}(x_n)\bar{\cal M}(y_1)\cdots   \bar{\cal M}(y_m)        \}                 \vert  0 \rangle \\
&=\left(\sum_{i=1}^n g_i \delta^{(3)} (x-x_i)- \sum_{j=1}^m   g_j \delta^{(3)} (x-y_i)\right)
\langle 0 \vert   T\{   {\cal M}(x_1)\cdots {\cal M}(x_n)\bar{\cal M}(y_1)\cdots  \bar{\cal M}(y_m)        \}                          \vert  0 \rangle
\, ,
\end{split}
\ee
where $g_i$ are magnetic monopole charges, and ${\cal M}$ are magnetic monopole operators that cannot be written locally in terms of the fields participating in the original CFT. Below, we will use the magnetic version of the Ward identity to derive the magnetic asymptotic symmetries of QED in flat space-time in the presence of magnetic monopoles.

\subsection{Weinberg soft theorems as symmetries of the S-matrix}
We turn now to a brief discussion on Weinberg soft theorems in Minkowski space-time.
Weinberg soft theorems can be seen to arise as a consequence of the asymptotic symmetries of QED. In this section we briefly review this statement. We follow closely \cite{Strominger:2015bla}. For more details, see \cite{Strominger:2017zoo,Kapec:2017tkm,Campiglia:2015qka}.

An infinite set of conserved charges can be constructed in Minkowski space-time. To see this, we start with the Maxwell equations in the presence of electric and magnetic charges,
\be
d*F=*j_E\, , \quad \text{and}\quad dF=*j_M\, .
\ee
The one-forms $j_E$ and $j_M$ are conserved in the sense that
\be
d*j_E=d^2*F=0\, , \quad \text{and}\quad d*j_M=d^2F=0\, .
\ee
More generally, we can define the following currents associated to any function $\varepsilon(x)$, which are also conserved
\be
*j^{\varepsilon}_E = d(\varepsilon *F)\, , \quad \text{and}\quad *j^{\varepsilon}_M = d(\varepsilon F)\, .
\ee
The charges associated to these currents read
\be\label{eq:charges}
\begin{split}
Q_E^{\varepsilon}(\Sigma)&=\int_{\Sigma} *j^{\varepsilon}_E  =\int_{\Sigma} \left(   d\varepsilon \wedge *F+\varepsilon *j_E    \right)\, , \\
Q_M^{\varepsilon}(\Sigma)&=\int_{\Sigma} *j^{\varepsilon}_M  =\int_{\Sigma} \left(   d\varepsilon \wedge F+\varepsilon *j_M   \right)\, .
\end{split}
\ee
The second terms in these expressions are weighted integrals of the electric/magnetic currents, and they will be refered to as the ``hard" terms. The first terms are linear in the electro-magnetic field and create photons. Conservation implies that the S-matrix operator must commute with these charges. We thus write
\be\label{eq:QScom}
\langle \text{out} \vert       Q_{E/M}^{\varepsilon}(\Sigma_+) {\cal S}-{\cal S}   Q_{E/M}^{\varepsilon}(\Sigma_-)        \vert \text{in}\rangle =0\, ,
\ee
where $\Sigma_{\pm}$ are late and early time Cauchy slices. Together with a few extra assumptions, this can be shown to be equivalent to Weinberg soft theorems. We will now discuss this equivalence briefly. 

For simplicity, we will focus on a theory with electrically charged particles. Of particular interest is the conservation law associated to parameters $\varepsilon(\hat{x})$ that  only depend on the angular coordinates. In this case, the photonic term in \eqref{eq:charges} involves photons with vanishing frequency, which are known as ``soft" photons. The soft part of the charges can be written as
\be\label{eq:SoftQ}
\begin{split}
Q_E^{\text{soft}}(\Sigma_+) &=  {-1\over 8\pi^2} \lim_{\omega_{\vec{q}}\rightarrow 0} \omega_{\vec{q}} \int d^2z \, \varepsilon(z,\bar{z}) \left[    \partial_{\bar{z}}\left(  {\sqrt{2}\over 1+z\bar{z}}  \sqrt{2\omega_{\vec{q}}}\, \hat{a}^{(-)}_{\vec{q}}\right) + \partial_{z}\left(  {\sqrt{2}\over 1+z\bar{z}}  \sqrt{2\omega_{\vec{q}}}\, \hat{a}^{(+)}_{\vec{q}}\right)               \right]\, ,
\end{split}
\ee
where the creation operators appearing here involve a soft momentum vector $\vec{q}$ with magnitude $\omega_{\vec{q}}\rightarrow 0$ pointing in the direction specified by $z,\bar{z}$. The charge at $\Sigma_-$ can be related to the one at $\Sigma_+$ through ${\cal CPT}$ invariance when inserted inside the S-matrix. We have
\be
\langle \text{out} \vert       Q_{E/M}^{\text{soft}}(\Sigma_+) {\cal S}\vert \text{in}\rangle =- \langle \text{out} \vert    {\cal S}   Q_{E/M}^{\text{soft}}(\Sigma_-)        \vert \text{in}\rangle \, .
\ee
This allows us to write the soft part of formula \eqref{eq:QScom} as a scattering amplitude involving charges only acting on outgoing states.

The hard part of the charges can also be written explicitly. We will consider currents associated to point-like charged particles, such that
\be
j_E^{\mu}(x)=\sum_i q_i \int ds \,p^{\mu}_i \delta^{(4)}(x-X_i(s))\, ,
\ee
where $i$ labels each particle, and $X_i(s)$ is the trajectory. Using this expression we can evaluate the hard part of the charge and obtain
\be\label{eq:HardQ}
\begin{split}
Q_E^{\text{hard}}(\Sigma_+)&=\lim_{t\rightarrow \infty} \sum_i q_i \,\varepsilon\left( x \right)\vert_{r={\vert \vec{p}\vert \over p^0} t, \hat{x}=\hat{p}_i }\, ,
\end{split}
\ee
and similarly for the early time Cauchy slice. Using formulas \eqref{eq:HardQ} and \eqref{eq:SoftQ} and plugging them in equation \eqref{eq:QScom} yields a formula relating scattering amplitudes involving soft photons with amplitudes involving exclusively hard particles. This is however, still not equivalent to Weinberg's soft theorem.

A further assumption on the Maxwell radiative data must now be made that makes the two terms appearing in formulas \eqref{eq:SoftQ} equal to each other. We need to require that magnetic monopoles are absent, i.e.
\be\label{eq:assumption}
F_{z\bar{z}}\vert_{{\cal I}^+_-}=F_{z\bar{z}}\vert_{{\cal I}^-_+} = 0\, .
\ee
Lastly, a particular choice of function $\varepsilon(x)$ yields Weinberg soft theorems. Focusing for simplicity on massless charged hard particles, the choices (one for each polarization)
\be\label{eq:varepsilonchoice}
\varepsilon(x)={1\over z-z'} \, , \quad \text{and}\quad \varepsilon(x)={1\over \bar{z}-\bar{z}'} \, ,
\ee
yield
\be\label{eq:WST}
\langle \text{out}\vert \sqrt{2\omega_{\vec{q}}} \, a^{(\pm)}_{\vec{q}} {\cal S} \vert \text{in}\rangle 
\sim  \left[   \sum_i  q_i {p_i\cdot \epsilon^{\pm}\over p_i\cdot q} -\sum_j   q_j {p_j\cdot \epsilon^{\pm}\over p_j\cdot q} \right] \langle \text{out}\vert {\cal S}\vert \text{in}\rangle\, .
\ee
 This result is well known as Weinberg's soft photon theorem \cite{Weinberg:1965nx}, and it relates a scattering event involving a soft photon to a scattering event without it.
 
 Even though the results presented here involve only massless hard particles, it is possible to obtain soft theorems involving massive matter by implementing a different choice of parameter $\varepsilon(x)$. This was explored in detail in \cite{Campiglia:2015qka,Kapec:2017tkm}. In this case, the parameter reads
 \be\label{eq:MinkepChoice}
 \varepsilon(x)=\int d^2\hat{x}'\, {1\over 4\pi }{ { t^2-r^2 }  \over {\left( t-r\hat{x}\cdot\hat{x}'  \right)^2 } } \varepsilon(\hat{x}')\, .
 \ee
where $\varepsilon(\hat{x}')$ are the same as in equation \eqref{eq:varepsilonchoice}. Under this choice, the second terms in the expression for the charges \eqref{eq:charges} now yield the hard terms in Weinberg soft theorems. The soft photon contributions are harder to deal with, but they can be shown to correspond to formulas \eqref{eq:SoftQ}. The final result is the same as \eqref{eq:WST} above, except that the momenta $p_i$ are also allowed to be time-like.

An alternative way of looking at the results presented here is that Weinberg soft theorems imply the symmetries of the S-matrix as written in \eqref{eq:QScom}. However, in order to derive the soft theorem from the statement about symmetries, one needs to make some extra assumptions like \eqref{eq:assumption}. In the following section we will show how the map between scattering states and conformal operators discussed in section \ref{sec:GaugeFields}, together with conformal Ward identities implies the symmetries \eqref{eq:QScom}.

\subsection{From CFT physics to soft theorems}\label{sec:WardSoftE}
We now would like to understand how the physics described in the previous subsection arise from a statement in conformal field theories. Our first step is to construct a soft photon scattering state as a CFT operator.  We have discussed how to create photon scattering states in section \ref{sec:GaugeFields}. We are now taking the soft limit $\omega_{\vec{q}}\rightarrow 0$. We thus write
\be
\begin{split}
\lim_{\omega_{\vec{q}}\rightarrow 0}\sqrt{2\omega_{\vec{q}}} \,   a_{\vec{q}}^{(-)}   =&\lim_{\omega_{\vec{q}}\rightarrow 0}\lim_{L \to \infty}
{-1\over 4\omega_q}
{1+z_q \bar{z}_q\over{\sqrt{2}}} 
 \int d\tau' \, e^{-i \omega_{\vec{q}}L\left(  {\pi \over 2}-\tau'  \right)}
\int d^2 z'   {1\over z'-z_q}    \partial_{z'} j^-_{\bar{z}'}(\tau',z',\bar{z}')    \, ,\\
\lim_{\omega_{\vec{q}}\rightarrow 0}\sqrt{2\omega_{\vec{q}}} \,   a_{\vec{q}}^{(+)}   =&\lim_{\omega_{\vec{q}}\rightarrow 0}\lim_{L \to \infty}
{-1\over 4\omega_q}
{1+z_q \bar{z}_q\over{\sqrt{2}}} 
 \int d\tau' \, e^{-i \omega_{\vec{q}}L\left(  {\pi \over 2}-\tau'  \right)}
\int d^2 z'   {1\over \bar{z}'-\bar{z}_q}    \partial_{\bar{z}'} j^-_{z'}(\tau',z',\bar{z}')  \, .
\end{split}
\ee
Note hat the large $L$ limit is taken before the soft limit. {In that limit the integrals over global time are dominated by operators evaluated at a window of size ${\cal O}(1/L)$ around $\tau'=\pi/2$. This follows from the same arguments presented in section \ref{sec:ScatteringStatesScalar} for scalar operators. The large phase pre-factor acquires a finite phase in this window of global time. The soft limit is implemented afterwards, setting that finite phase to zero. This sets the phase in the global time integral to one.} We can thus simply integrate the current insertions over the small windows of global time at $\tau=\pm{\pi\over 2}$ which we denote $\tilde{\cal I}^{\pm}$, and obtain
\be
\begin{split}
\lim_{\omega_{\vec{q}}\rightarrow 0}\sqrt{2\omega_{\vec{q}}} \,   a_{\vec{q}}^{(-)}   =&\lim_{\omega_{\vec{q}}\rightarrow 0}
{-1\over 4\omega_q}
{1+z_q \bar{z}_q\over{\sqrt{2}}} 
\int_{\tilde{\cal I}^{\pm}}d^3 x' {1\over z'-z_q}    D^{\bar{z}'} j^-_{\bar{z}'}(\tau',z',\bar{z}')    \, ,
\end{split}
\ee
and similarly for annihilation operators. Scattering amplitudes involving soft photon modes are then simply CFT correlators involving insertions of the conserved current. These are constrained by the CFT Ward identities written above in formula \eqref{eq:EQ1IBP}, which we repeat here
\be\label{eq:EQ1v2}
\begin{split}
&\int d^3 x\,\alpha(x)  \partial_{\mu} \langle 0 \vert     T\{   j^{\mu}(x)  X     \}                 \vert  0 \rangle =\left(\sum_{i=1}^n q_i \alpha (x_i)- \sum_{j=1}^m   q_j \alpha (y_j) \right)
\langle 0 \vert    T\{    X      \}                 \vert  0 \rangle
\, .
\end{split}
\ee
{Here, we integrate the Ward identity over a specific region of the conformal theory. This region consists of the AdS boundary in Lorentzian signature between the past of $\tilde{\cal I}^-$ to the future of $\tilde{\cal I}^+$, together with two Euclidean caps $\partial{\cal M}_{\pm}$, resulting in a pill-shaped region like the one shown in figure \ref{fig:Interpretation}.} In equation \eqref{eq:EQ1v2}, $X$ stands for all insertions of primary operators charged under the $U(1)$ current. Explicitly, we have
\be
X=\prod_i  {\cal O}_i(x_i)\prod_j  \bar{\cal O}_i(y_j)\, .
\ee
By smearing the locations of the primary operators we can turn them into creation/annihilation operators involving hard particles in Minkowski space, as shown in sections \ref{sec:MasslessRec} and \ref{sec:MassiveRec}. In the massless case, operators will be light and be placed in the regions $\tilde{\cal I}^{\pm}$, while in the massive case, the operators will be heavy and will live on the Euclidean half-spheres $\partial{\cal M}_{\pm}$. There is a particular choice of parameter $\alpha(x)$ that turns the right hand side of \eqref{eq:EQ1v2} into the hard terms in Weinberg's soft theorem. Namely,
\be\label{eq:alphaCFT}
\alpha(x)=\lim_{\rho\rightarrow{\pi \over 2}}\int d^2 \hat{x}' \, {1\over 4\pi}{{\cos^2\rho-\cos^2\tau }\over{\left(   \sin\tau -\sin\rho\, \hat{x}\cdot\hat{x}'  \right)^2}} \varepsilon(\hat{x}')\, .
\ee
This formula is derived in appendix \ref{app:alpha}, and it obeys the property that at $\tilde{\cal I}^{\pm}$ there is no time dependence, such that
\be
\alpha(x)\vert_{\tilde{\cal I}^{\pm}}=\varepsilon(\hat{x})\, .
\ee
This choice of parameter $\alpha(x)$, together with the choices \eqref{eq:varepsilonchoice} for $\varepsilon(\hat{x})$ turn the right hand side of  \eqref{eq:EQ1v2} into the hard part of Weinberg's soft theorem. This includes both massless and massive particles. We now need to show that an appropriate smearing of the left hand side of \eqref{eq:EQ1v2} corresponds to the Minkowski soft charges discussed in the previous section. The integral over the boundary receives contributions only from $\tilde{\cal I}^{\pm}$ and the  Euclidean half-spheres $\partial{\cal M}_{\pm}$.  All the primary operators in the correlator are located strictly on these regions. Integrating by parts, and using the fact that $\alpha(x)$ is time independent for the contributions at $\tilde{\cal I}^{\pm}$ , we can write
\be\label{eq:ArgumentsHolographic}
\begin{split}
\int d^3 x\, \partial_{\mu}\alpha(x) \langle 0 \vert     T\{   j^{\mu}(x)  X     \}                 \vert  0 \rangle =\int_{\partial{\cal M}_{\pm}} d^3 x\, \partial_{\mu}\alpha(x) \,  \langle 0 \vert     T\{   j^{\mu}(x)  X     \}                 \vert  0 \rangle  \\
- \int_{\tilde{\cal I}^{\pm}}  d^3 x\,\varepsilon(\hat{x}) \left[   D^z \langle 0 \vert     T\{   j_z(x)  X     \}                 \vert  0 \rangle +  D^{\bar{z}} \langle 0 \vert     T\{   j_{\bar{z}}(x)  X     \}                 \vert  0 \rangle \right]\, .
\end{split}
\ee
The terms in the integral over Euclidean regions $\partial{\cal M}_{\pm}$ involve only the Coulombic part of the current and can be shown to vanish for the choice of $\alpha(x)$ written above in equation \eqref{eq:alphaCFT}\footnote{
This could be checked explicitly in the CFT using the formula for the Coulombic part of the current written in section \ref{sec:Coulomb}. Such a check is actually non-trivial, as it involves the computation of multi-point correlators involving the double trace operators appearing in the definition of $j_{\mu}^{\text{C}}$. The argument can also be made simpler in the bulk by computing AdS Li\'enard-Wiechert potentials, which compute expectation values of the Coulombic part of the current when evaluated at the boundary of AdS. For a sample calculation, see appendix \ref{app:LWAdS}.
}. The second line can also be shown to receive only contributions from the radiative part of the current. We thus conclude that the right hand side of the Ward identity reads
\be\label{eq:auxSP1}
\begin{split}
\int d^3 x\, \alpha(x) \partial_{\mu}\langle 0 \vert     T\{   j^{\mu}(x)  X     \}                 \vert  0 \rangle &=\\
 \int_{\tilde{\cal I}^{\pm}}  d^3 x\,\varepsilon(\hat{x}) &\left[   D^z \langle 0 \vert     T\{   j^{(\text{R})}_z(x)  X     \}                 \vert  0 \rangle  +  D^{\bar{z}} \langle 0 \vert     T\{   j^{(\text{R})}_{\bar{z}}(x)  X     \}                 \vert  0 \rangle \right]\, .
\end{split}
\ee
The current operators appearing here correspond to the following soft photon operators (ignoring the radiative label)
\be\label{eq:auxSP2}
\begin{split}
 \int_{\tilde{\cal I}^{+}} d^3x \,  \varepsilon(\hat{x})
D^{\bar{z}} j^-_{\bar{z}}   =& \lim_{\omega_{\vec{q}}\rightarrow 0}{2\over \pi}\omega_{\vec{q}} \, \int d^2z_q \, \varepsilon(\hat{q}) 
\, \partial_{\bar{z}_q}\left(   {\sqrt{2}\over 1+z\bar{z}} \sqrt{2\omega_{\vec{q}}}\, a^{(-)}_{\vec{q}}  \right)   \, , \\
 \int_{\tilde{\cal I}^{+}} d^3x \,  \varepsilon(\hat{x})
 D^{z} j^-_{z}  =&\lim_{\omega_{\vec{q}}\rightarrow 0}{2\over \pi}\omega_{\vec{q}}\, \int d^2z_q \, \varepsilon(\hat{q}) \, 
 \partial_{z_q}\left(   {\sqrt{2}\over 1+z\bar{z}} \sqrt{2\omega_{\vec{q}}}\, a^{(+)}_{\vec{q}}  \right)   \, .
\end{split}
\ee
This implies that the left hand side of the Ward identity is precisely the soft part of the charge in Minkowski space at ${\cal I}^+_-$, which was written explicitly in equation \eqref{eq:SoftQ}. Here, we have only written explicitly the contributions from $\tilde{\cal I}^+$, but similar expressions can be written for $\tilde{\cal I}^-$, which yield the soft charges at ${\cal I}^-_+$.  { Note that equation \eqref{eq:auxSP1} contains terms which involve both positive and negative frequency modes. Equation \eqref{eq:auxSP2} only considers the contribution to  $\tilde{\cal I}^+$. Since only a single soft photon is present in \eqref{eq:auxSP1}, its positive frequency contribution turns into a creation operator which annihilates the vaccum-bra vector. Thus only the negative frequency part contributes. An analogous story ensures that at $\tilde{\cal I}^-$ only the positive frequency modes contribute.  }

To put all the pieces together, we have concluded that the left hand side of the Ward identity \eqref{eq:EQ1v2} turns into the insertions of the soft charge operators into the S-matrix.   We have also shown that the right hand side of the Ward identity turns into the insertions of the hard charge operators.  This concludes the proof that conformal Ward identities in CFT$_3$ imply the invariance of the Minkowski S-matrix under the charges \eqref{eq:charges}. One can further proceed as in the previous section to show this implies Weinberg's soft theorems.

\subsection{From CFT physics to magnetic soft theorems }\label{sec:MST}
In the previous section, we have analyzed how the Ward identity in a conformal field theory dual to an AdS theory with the standard boundary conditions leads to the invariance of the S-matrix under the electric charges \eqref{eq:charges}. Here, we analyze AdS theories with alternative boundary conditions. 

In section \ref{sec:FLE} (and appendix \ref{MaxfieldReconstructionElectric}), we have discussed the reconstruction of photon scattering states as operators in a conformal field theory dual to an AdS theory with ``electric'' boundary conditions. As a reminder, such conditions fix the electric field at the boundary, but allow the magnetic field created by magnetic monopoles in the bulk theory. The result of the analysis is the creation/annihilation operators in equation \eqref{eq:ElectricBCphotons}, which we repeat here.
\be
\begin{split}
\sqrt{2\omega_{\vec{q}}} \,   v_{\vec{q}}^{(-)}=&{1\over 4\omega_q}{1+z_q \bar{z}_q\over{\sqrt{2}}}  \int d\tau' \,  e^{-i \omega_{\vec{q}} L\left({\pi\over 2}-\tau' \right)}
\partial_{\tau'}A^-_{z'}(\tau',z_q',\bar{z}_q')
\, , \\
\sqrt{2\omega_{\vec{q}}} \,  v_{\vec{q}}^{(+)}=&{1\over 4\omega_q}{1+z_q \bar{z}_q\over{\sqrt{2}}}  \int d\tau' \,  e^{-i \omega_{\vec{q}} L\left({\pi\over 2}-\tau' \right)}
\partial_{\tau'}A^-_{\bar{z}'}(\tau',z_q',\bar{z}_q')\, .
\end{split}
\ee
In this section, we will be interested in the soft limit of these operators, such that $\omega_{\vec{q}}\rightarrow 0$. Like in the previous section, the soft limit is taken after the large $L$ limit, so the CFT operators are always integrated over the CFT regions $\tilde{\cal I}^{\pm}$. The CFT operators $A_{\mu}$ are dynamical gauge fields, whose dual field strength obeys a broken Bianchi identity. This was discussed in equation \eqref{eq:EQ2}, which we repeat here
\be\label{eq:EQ2Repeat}
\begin{split}
&\partial_{\mu} \langle 0 \vert    T\{  (*f)^{\mu}(x) X     \}                 \vert  0 \rangle =\left(\sum_{i=1}^n g_i \delta^{(3)}(x-x_i)-\sum_{j=1}^m  g_j  \delta^{(3)}(x-y_j) \right)
\langle 0 \vert   T\{  X     \}                          \vert  0 \rangle
\, .
\end{split}
\ee
Here,  $X$ stands for insertions of CFT operators with magnetic charge. We will follow a very similar strategy as in the previous section. We first integrate the Ward identity over the locations of the operators in X such that they become asymptotic creation/annihilation operators in Minkowski space. Choosing $\alpha(x)$ as in formula \eqref{eq:alphaCFT} makes the right hand side of the Bianchi identity \eqref{eq:EQ2} turn into the hard part of Weinberg's soft theorem, but with magnetic charges instead of electric ones. 

The treatment of the left hand side of \eqref{eq:EQ2Repeat} proceeds the same way as in the previous section, which leads to
\be
\begin{split}
\int d^3 x\, &\alpha(x)\, \partial_{\mu} \langle 0 \vert     T\{   (*f)^{\mu}(x)  X     \}                 \vert  0 \rangle = \\
& \int_{\tilde{\cal I}^{\pm}}  d^3 x\,\varepsilon(\hat{x}) \left[   D^z \langle 0 \vert     T\{   (*f)^{(\text{R})}_z(x)  X     \}                 \vert  0 \rangle +  D^{\bar{z}} \langle 0 \vert     T\{   (*f)^{(\text{R})}_{\bar{z}}(x)  X     \}                 \vert  0 \rangle \right]\, .
\end{split}
\ee
Where again, only the radiative part of the current appears. This can be written explicitly in terms of the CFT gauge field operator $A^{(\text{R})}_{\mu}$. Dropping the radiative label, we have
\be
\begin{split}
\int d^3 x\, &\alpha(x)\, \partial_{\mu} \langle 0 \vert     T\{   (*f)^{\mu}(x)  X     \}                 \vert  0 \rangle = \\
&2 i \int_{\tilde{\cal I}^{\pm}}  d\tau d^2z \,\varepsilon(\hat{x}) \left[   \partial_{\bar{z}} \langle 0 \vert     T\{  \partial_{\tau}A_z(x)  X     \}                 \vert  0 \rangle -\partial_z \langle 0 \vert     T\{   \partial_{\tau}A_{\bar{z}}(x)  X     \}                 \vert  0 \rangle \right]\, .
\end{split}
\ee
The operators appearing in this expression can be written in terms of soft photon creation operators as follows. For the term involving $\tilde{\cal I}^+$, we have
\be
\begin{split}
 \int_{\tilde{\cal I}^+} d\tau  d^2z \, \varepsilon(\hat{x})\, 
  \partial_{\bar{z}}    \partial_{\tau}A^-_{z}(\tau,z,\bar{z})=&
      \lim_{\omega_{\vec{q}}\rightarrow 0}
      4\omega_{\vec{q}}
      \int d^2 z_q \, \varepsilon(\hat{q})
    \partial_{\bar{z}_q}\left(         {{\sqrt{2}}\over{1+z \bar{z}}} 
\sqrt{2\omega_{\vec{q}}} \,   v_{\vec{q}}^{(-)} \right)
\, , \\
 \int_{\tilde{\cal I}^+} d\tau  d^2z \, \varepsilon(\hat{x})\, 
  \partial_{z}    \partial_{\tau}A^-_{\bar{z}}(\tau,z,\bar{z})=&
      \lim_{\omega_{\vec{q}}\rightarrow 0}
      4\omega_{\vec{q}}
      \int d^2 z_q \, \varepsilon(\hat{q})
    \partial_{z_q}\left(         {{\sqrt{2}}\over{1+z \bar{z}}} 
\sqrt{2\omega_{\vec{q}}} \,   v_{\vec{q}}^{(+)} \right)
\, .
\end{split}
\ee
This implies
\be
\begin{split}
- 2 i \int_{\tilde{\cal I}^{+}}  d\tau d^2z \,\varepsilon(\hat{x}) \left[   \partial_{\bar{z}} \langle 0 \vert     T\{  \partial_{\tau}A_z(x)  X     \}                 \vert  0 \rangle -\partial_z \langle 0 \vert     T\{   \partial_{\tau}A_{\bar{z}}(x)  X     \}                 \vert  0 \rangle \right] = \langle 0 \vert     T\{    Q_M^{\text{soft}}(\Sigma_+) X     \}                 \vert  0 \rangle \, .
\end{split}
\ee
where we have defined the operator
\be
Q_M^{\text{soft}} (\Sigma_+) =  - 8 i \lim_{\omega_{\vec{q}}\rightarrow 0} \omega_{\vec{q}}
      \int d^2 z_q \, \varepsilon(\hat{q})
      \left[ 
       \partial_{\bar{z}_q}\left(         {{\sqrt{2}}\over{1+z \bar{z}}} 
\sqrt{2\omega_{\vec{q}}} \,   v_{\vec{q}}^{(-)} \right)
-
 \partial_{z_q}\left(         {{\sqrt{2}}\over{1+z \bar{z}}} 
\sqrt{2\omega_{\vec{q}}} \,   v_{\vec{q}}^{(+)} \right) 
      \right] \, .
\ee
similar expressions can be written for the terms coming from the CFT region $\tilde{\cal I}^-$, and they involve an operator $Q_M^{\text{soft}} (\Sigma_-)$ constructed out of creation and annihilation operators in the asymptotic past. These are precisely expressions for magnetic soft charges in Minkowski space. With this, we conclude that the broken Bianchi identity \eqref{eq:EQ2Repeat} implies the asymptotic magnetic symmetries of a flat space theory with magnetic monopoles, upon the implementation of the map between creation/annihilation operators in Minkowski space and conformal operators written in \eqref{eq:ElectricBCphotons}.

\section{Conclusions}\label{sec:conc}
In this paper we have developed a method to construct CFT operators whose correlation functions yield flat space S-matrix elements. This method was then applied to obtain CFT operators dual to photon creation and annihilation operators. We could then reproduce well-known features of the S-matrix -- Weinberg soft theorems -- within the framework of flat-space holography. It has been argued that Weinberg soft theorems arise as the Ward identities of asymptotic gauge transformations, also called large gauge transformations. Here, we have demonstrated that they can also be thought of arising from CFT Ward identities in the flat-space limit. 

Given the methods presented in this paper, a natural next step would be to derive an expression for fermion, non-abelian gauge field and graviton creation/annihilation operators. The latter will take the form of a smeared energy-momentum tensor, again localized at $\tau = \pm \frac \pi 2$ for outgoing and incoming states, respectively. Obtaining results for fermion, as well as non-Abelian gauge field scattering states would be a crucial step in reproducing phenomenologically interesting scattering cross-sections from flat-space holography. Although it is not obvious that calculating scattering amplitudes as CFT correlators would be any easier that modern perturbative methods, it allows to apply new techniques, like those of the bootstrap program \cite{Paulos:2016fap}, to scattering cross-sections.

The flat limit of AdS/CFT might reveal more details about the infrared structure of flat space-time. As an example, note that a crucial step in our analysis was to assume that the energy as measured by a CFT observer is of order $\omega_\text{flat} L$ and thus diverges in the flat space limit. To obtain the soft theorems the energy is taken to vanish after the large $L$ limit is taken. This raises a question about the low-energy modes with CFT energies of order $\mathcal O(1)$. For those modes the sum over quantum numbers $\kappa$ does not turn into a continuum and they should appear as IR modes in the bulk theory. It would be interesting to understand if those modes are frozen or if there is dynamics associated with them. If the latter case is true, those modes might be a mechanism for how information can be stored in the infrared in flat space. We hope to be able to report results of this analysis in the near future.

Concerning the IR structure of flat space physics, it is well known that infrared divergences trivialize the Fock basis S-matrix. The existence of these problematic divergences can be tracked back to the assumption of asymptotic decoupling. The introduction of Faddeev-Kulish states \cite{Kulish:1970ut} solves this issue by dressing the scattering states with soft photons, in such a way that the IR divergences of Weinberg soft theorems cancel the divergences of the S-matrix, see also \cite{Carney:2018ygh,Neuenfeld:2018fdw}. While AdS/CFT acts as a natural IR regulator, IR divergences will show up once the flat limit is taken. AdS/CFT suggests a solution related to that of Faddeev-Kulish, by relaxing the condition of asymptotic decoupling. This could be done by making use of the HKLL technology to reconstruct bulk operators that interact with the long wavelength modes of the electro-magnetic field.

From the point of view of the CFT, flat space scattering states are a special class of states which behave as if they were free for most of the CFT time-evolution. Equivalently, we have found a set of non-local CFT operators which obey a simple creation/annihilation operator algebra. From the point of view of AdS/CFT and in particular HKLL bulk reconstruction, the existence of such operators is not very surprising and can be explained in terms of error-correcting codes \cite{Almheiri:2014lwa}. For a different construction of CFT operators obeying canonical commutation relations, see \cite{Papadodimas:2012aq,Magan:2020iac}. Light ray operators \cite{Kravchuk:2018htv} are another class of non-local operators which again obey a simple but non-trivial algebra and it was shown in \cite{Cordova:2018ygx} that in a Lorentz invariant CFT, certain operators reproduce the BMS algebra. It would be interesting to explore whether these operators are related to the ones discussed throughout this work. 

There have been several intriguing observations in the literature which hint at the possibility at a possibly exotic CFT might underlie flat space scattering, in the sense that the calculation of scattering amplitudes can be understood in terms of correlation functions of this CFT. We believe that this paper gives further insights into the structure of a dual theory of flat space scattering. However, taking into account the absence of an explicit construction of flat holography (which is available in AdS/CFT), one should perhaps understand the above results as a toolbox to explore the qualitative features a putative theory dual to flat space must have, rather than a bona-fide derivation of flat-space holography from AdS/CFT.

\acknowledgments
It is a pleasure to thank Felix Haehl, Dan Kapec, Sabrina Pasterski, Silviu Pufu and Antony Speranza for useful conversations.
EH acknowledges support from the Gravity Initiative at Princeton University.
DN acknowledges support from the Simons Foundation through the “It from Qubit” collaboration.
Research at Perimeter Institute is supported in part by the Government of Canada through the Department of Innovation, Science and Economic Development Canada and by the Province of Ontario through the Ministry of Colleges and Universities. This research was supported in part by the National Science Foundation under Grant No. PHY-1748958. We also thank ICTP-SAIFR for hospitality during the S-matrix bootstrap IV workshop organized by the Simons Collaboration on the Non-perturbative Bootstrap.
\appendix
\addtocontents{toc}{\protect\setcounter{tocdepth}{1}}

\section{Complex coordinates on $S^2$}\label{app:coordS2}
The standard parametrization of a two-sphere is done with angles $\theta\in[0,\pi]$ and $\phi\in(0,2\pi]$. Alternatively, one can define complex coordinates $z, \bar{z}$ such that
\be
\cos\theta={1-z\bar{z}\over 1+z\bar{z}}\, , \quad  \sin\theta\cos\phi={z+\bar{z}\over1+z\bar{z}}\, , \quad \sin\theta\sin\phi=-i{z-\bar{z}\over 1+z\bar{z}}\, .
\ee
The metric in these coordinates is
\be
ds^2={4 dz d\bar z\over(1+z\bar{z})^2}\, .
\ee
The Laplace operator eigenvalue equation obeyed by spherical harmonics,
\be
\left(\partial_{\theta}^2+{1\over \sin^2\theta} \partial_{\phi}^2 +\cot\theta \partial_{\theta}\right)Y_l^m(\hat{\Omega}) = -l (l+1)Y_l^m(\hat{\Omega}),
\ee
takes a different form in the new coordinates,
\be
(1+z\bar{z})^2 \partial \bar{\partial}Y_l^m(z,\bar{z}) = -l (l+1)Y_l^m(z,\bar{z}).
\ee
In complex coordinates, the Dirac delta function reads 
\be
\delta^{(2)}(\hat{\Omega},\hat{\Omega}') = \delta(\phi-\phi')\delta(\cos\theta-\cos\theta')=  {1 \over 2}(1+z \bar{z})^2 \delta^{(2)}(z,z')\, ,
\ee
The Greens fuction associated to the Laplace operator must obey the differential equation
\be
(1+z\bar{z})^2 \partial \bar{\partial} G(z,\bar{z})={1\over 2}(1+z \bar{z})^2 \delta^{(2)}(z,z') - \frac{1}{4\pi}\, ,
\ee
which is solved by 
\be\label{eq:GS2}
G={1\over 4\pi i}\log |z-z'|^2 - \frac{1}{4\pi} \log(1 + z \bar z)- \frac{1}{4\pi} \log(1 + z' \bar z') \, .
\ee
In complex coordinates, the derivative operators read
\be
\sin\theta\partial_{\theta}= z\partial+\bar{z}\bar{\partial}\, , \quad \partial_{\phi}= i (z\partial-\bar{z}\bar{\partial})\, .
\ee

\section{Scalar HKLL operators}\label{app:ScalarHKLL}
This section reviews operator reconstruction according to HKLL and derives the formulas used in the main text. Long before and after the scattering, scalar fields obey the Klein-Gordon equation. Normalizable solutions in AdS$_4$, which are the ones we are interested in, are linear combinations of modes of the form
\be
\phi^{\pm}_{\omega,l,m}=\frac{1}{\mathcal N_{\Delta, \omega, l}}e^{\pm i \omega \tau} Y_l^{m}(\Omega)^{(*)} \sin^l \rho \cos^{\Delta}\rho \, {}_2F_1\left( {\Delta+l-\omega\over 2},{\Delta+l+\omega\over 2};\Delta-{1\over 2} \Big| \cos^2\rho   \right)\, ,
\ee
where the superscript ${}^{(*)}$ of the spherical harmonic $Y_l^m(\Omega)$ indicates that we should take the complex conjugate for $\phi^-_{\omega, l, m}$.
Those modes are only regular in the center of AdS ($\rho = 0$), if
\be
\omega_{\kappa, l}= \Delta+l+2\kappa\, , \quad \text{with}\quad \kappa\in \mathbb{Z}^+\, .
\ee
Normalization with respect the the Klein-Gordon inner product require that the normalization constant is
\be
\mathcal N_{\Delta, \kappa, l} = L \sqrt{\frac{\kappa! \Gamma(\Delta - \frac 1 2)^2 \Gamma(\kappa + l+ \frac 3 2)}{\Gamma(\Delta+\kappa+l)\Gamma(\Delta + \kappa - \frac 1 2) }}\, .
\ee
We will denote regular, normalized modes with $\phi^{\pm}_{\kappa,l,m}$. General solutions of the free equations of motion can thus be written as
\be\label{eq:PhiFreeSol}
\phi(\rho,x)=\sum_{\kappa\in\mathbb{Z}^+}\sum_{l,m} \left( a^{+}_{\kappa,l,m}\phi^{+}_{\kappa,l,m} + a^{-}_{\kappa,l,m}\phi^{-}_{\kappa,l,m}\right) \, .
\ee
The modes $\phi^{\pm}_{\kappa,l,m}$ satisfy orthonormality relations, such that formula \eqref{eq:PhiFreeSol} can be promoted to an operator statement through canonical quantization,
\be\label{eq:PhiFreeOp}
\hat{\phi}(\rho,x)=\sum_{\kappa\in\mathbb{Z}^+}\sum_{l,m} \left( \hat{a}^{+}_{\kappa,l,m}\phi^{+}_{\kappa,l,m} + \hat{a}^{-}_{\kappa,l,m}\phi^{-}_{\kappa,l,m} \right)\, .
\ee
The dual operator at the boundary is then
\be
\begin{split}
{\cal O}(x)& =\sqrt{\frac{4 \pi \Gamma(\frac 3 2) \Gamma(\Delta - \frac 1 2)}{\Gamma(\Delta)}} \lim_{\rho\rightarrow{\pi \over 2}} \cos^{-\Delta} \rho \, \hat{\phi}(\rho,x) \\
& = \frac 1 {\mathcal N_{\mathcal O, \kappa,l}} \sum_{\kappa\in\mathbb{Z}^+}\sum_{l,m} \left( \hat{a}^{+}_{\kappa,l,m}e^{i \omega_{\kappa,l}\tau} Y_l^m(\Omega)
+ \hat{a}^{-}_{\kappa,l,m}e^{-i \omega_{\kappa,l}\tau} Y_l^{m}(\Omega)^{(*)} \right) \, ,
\end{split}
\ee
with
\be
\mathcal N_{\mathcal O, \kappa,l} = \sqrt{\frac{\Gamma(\kappa + 1)\Gamma(\Delta)\Gamma(\Delta - \frac 1 2) \Gamma(\kappa+l + \frac 3 2)}{4 \pi \Gamma(\frac 3 2) \Gamma(\Delta + \kappa + l) \Gamma(\Delta + \kappa - \frac 1 2)}}.
\ee
The normalizing factor $N_{\mathcal O, \kappa,l}$ ensures that the CFT two-point function is canonically normalized.

The objective now is to write $\hat{a}^{\pm}_{\kappa,l,m}$ in terms of the CFT operator ${\cal O}$ and plug the solution back into \eqref{eq:PhiFreeOp} to obtain a representation of the bulk operator at the boundary. For this we define positive and negative frequency parts of the boundary operator
\be
{\cal O}^{\pm}(x)\equiv \frac 1 {\mathcal N_{\mathcal O, \kappa,l}}  \sum_{\kappa\in \mathbb{Z}^+}\sum_{l,m} \hat{a}^{\pm}_{\kappa,l,m}e^{\pm i \omega_{\kappa,l}\tau} Y_l^{m}(\Omega)^{(*)}\, .
\ee
We can invert this formula to obtain
\be
\begin{split}
\hat{a}^{\pm}_{\text{in},\kappa,l,m}&={\mathcal N_{\mathcal O, \kappa,l} \over \pi}\int_{- \pi}^{0}d\tau'\, \int d^2\Omega'\, {\cal O}^{\pm}(x') \, e^{\mp i \omega_{\kappa,l}\tau'}Y_l^m(\Omega')^{(\overline{*})}\, ,\\
\hat{a}^{\pm}_{\text{out},\kappa,l,m}&={\mathcal N_{\mathcal O, \kappa,l} \over \pi}\int_{0}^{\pi}d\tau'\, \int d^2\Omega'\, {\cal O}^{\pm}(x') \, e^{\mp i \omega_{\kappa,l}\tau'}Y_l^m(\Omega')^{(\overline{*})}\, .
\end{split}
\ee
The superscript ${}^{(\overline{*})}$ instructs to use complex conjugation if we choose $\hat a^+_{\kappa, l, m}$.
Plugging these expressions back into \eqref{eq:PhiFreeOp} naively yields,
\be
\begin{split}
\hat{\phi}(\rho,x)={ \mathcal N_{\mathcal O, \kappa,l} \over \pi}\int_{\mathcal T}d\tau'\, \int d^2\Omega'\,  
& \left( {\cal O}^{+}(x') \sum_{\kappa\in\mathbb{Z}^+}\sum_{l,m} \, e^{- i \omega_{\kappa,l}\tau'}Y_l^m(\Omega')  \phi^{+}_{\kappa,l,m} \right. \\
& +  \left. {\cal O}^{-}(x') \sum_{\kappa\in\mathbb{Z}^+}\sum_{l,m} \, e^{ i \omega_{\kappa,l}\tau'}Y_l^m(\Omega')^* \phi^{-}_{\kappa,l,m} \right) \, .
\end{split}
\ee
The integration region $\mathcal T$ for the time integral is between $-\pi$ and $0$ for in-fields and between $0$ and $\pi$ for out-fields. When the holographic coordinate is taken to the boundary, the kernels do not result in  delta functions so that one would recover the boundary operator. This can be resolved easily by extending the sum over $\kappa$ to all integers $\mathbb{Z}$. This can be done without adding extra terms, by noting that the added modes integrate to zero against ${\cal O}^+$ or ${\cal O}^-$. We conclude that
\be
\hat{\phi}(\rho,x)= \int_{\mathcal T} d\tau'\, \int d^2 \Omega\, \left[   K_{+}(\rho,x;x') {\cal O}^+(x')+ K_{-}(\rho,x;x') {\cal O}^-(x')        \right]\, ,
\ee
where the kernel $K_{+} = K_{-}^*$ reads
\be
\begin{split}
K_{+}(\rho,x;x')=& \frac 1 {\mathcal N_{K,\Delta}} \sum_{\kappa\in \mathbb{Z}}\sum_{l=0}^{\infty}\sum_{m=-l}^{m=l} e^{i \omega_{\kappa,l}(\tau-\tau')} Y_l^m(\Omega)Y_l^m(\Omega')^* \sin^l\rho \cos^{\Delta}\rho \,\\
&\times {}_2F_1\left(  -\kappa,l+\Delta+\kappa;\Delta-{1\over 2} \Big \vert \cos^2\rho  \right)\, .
\end{split}
\ee
The normalization ensures that a canonically normalized operator gives canonically normalized creation/annihilation operators in the flat space limit,
\be\label{eq:ScalarNorma}
\frac 1 {\mathcal N_{K,\Delta}} = \frac{1}{L}\sqrt{\frac{\Gamma(\Delta)}{4 \pi^3 \Gamma(\Delta - \frac 1 2)\Gamma(\frac 3 2)}}.
\ee

\section{Reconstruction of $U(1)$ gauge fields in global AdS$_4$}\label{app:MaxfieldReconstruction}
\subsection{Reconstruction with ``magnetic'' boundary conditions}\label{MaxfieldReconstructionMagnetic}
In this appendix we reconstruct local Maxwell fields in global AdS in terms of their boundary values at the boundary of Anti-de Sitter space. We will first consider AdS/CFT with magnetic boundary conditions, such that the dynamical part of the gauge field asymptotes to a CFT current as
\be\label{eq:MagneticBC}
{\cal A}_{\mu}(\rho,x)\xrightarrow[\rho \to \frac \pi 2]{} \cos\rho \, j_{\mu}(x)\, .
\ee
This defines the normalization for the current. In order to have a current two-point function with a particular normalization, an appropriate constant normalization factor has to be added to \eqref{eq:MagneticBC}.
In this appendix we will consider a bulk gauge field in Lorenz gauge. Gauge fixed fields can be considered physical observables, but they will have commutators that are local only to the extent allowed by the gauge constraint (see \cite{Heemskerk:2012np} for a nice discussion). Solutions to the Maxwell field equations with the boundary conditions \eqref{eq:MagneticBC} in Lorenz gauge can be constructed from a set of three different modes
\be\label{eq:Asols}
\begin{split}
{\cal A}^{V,l,m}_{\mu} =& L^{-2}\epsilon_{\tau\rho\mu}^{\quad\;\;\nu} e^{\pm i\omega \tau} \cos^{2}\rho \,R_V(\rho) \, \nabla_{\nu} Y_l^m(\hat{\Omega}) \, , \\
{\cal A}^{S,l,m}_{\mu} =& L^{-2}(l(l+1)-\omega^2)\epsilon_{\theta\phi\mu}^{\quad\;\;\nu} \sin^{-1}\theta \,\cot^2\rho\, Y_l^m(\hat{\Omega})\, \nabla_{\nu}(e^{\pm i\omega \tau} R_S(\rho))  \, ,\\
{\cal A}^{G,l,m}_{\mu} =& L^{-2}\nabla_{\mu}\left[ e^{\pm i\omega \tau}Q(\rho) Y_l^m(\hat{\Omega})   \right]   \, .
\end{split}
\ee
Here, $\epsilon$ is the Levi-Civita tensor with respect to the full metric. The indices $V$, $S$, and $G$ stand for ``vector" and ``scalar"-type degrees of freedom, as well as residual ``gauge". The naming convention is motivated by the transformation behavior under $SO(2)$ \cite{Ishibashi:2004wx}. We will completely ignore the pure gauge modes in what follows. The scalar-type solution given here is not in Lorenz gauge, but can be brought to the required form by a gauge transformation,
\be
\label{eq:gaugeToLorenz}
{\cal A}^{S,l,m}_{\mu} \to {\cal A}^{S,l,m}_{\mu}  + \nabla_\mu \chi \quad \text{with} \quad \chi = e^{\pm i\omega \tau} \partial_\rho(R_S(\rho)) Y_l^m(\hat{\Omega})\,.
\ee
The radial functions appearing in the above ansatz obey a hypergeometric differential equation
\be
R_{V/S}'' + \left(  \omega^2-{{l(l+1)}\over{\sin^2\rho}}   \right) R_{V/S}= 0  \, .
\ee
The solutions with the right fall-off at the boundary are
\be
\begin{split}
R_V(\rho)=& \sin^{l+1}\rho\, \cos\rho \, {}_2 F_1\left(  {2+l-\omega\over 2}, {2+l+\omega\over 2};{3\over 2}\Big\vert \cos^2\rho   \right) \, , \\
R_S(\rho)=& \sin^{l+1}\rho\, {}_2 F_1\left(  {1+l-\omega\over 2}, {1+l+\omega\over 2};{1\over 2}\Big\vert \cos^2\rho   \right) \, .
\end{split}
\ee
These functions are well behaved when expanded around the boundary, but present singular behavior when expanded around the center of AdS. In the asymptotic regions away from the scattering regions, this is unacceptable, since we should be able to trust the free field expansion. Like in the scalar case, the issue can be fixed by quantizing the frequency $\omega$. For the case of the vector-type degree of freedom, we must have
\be
\omega_V=2+l+2\kappa\, , \quad \text{with}\quad \kappa\in \mathbb{Z}^+\, ,
\ee
while for the scalar-type we must choose
\be
\omega_S=1+l+2\kappa\, , \quad \text{with}\quad \kappa\in \mathbb{Z}^+\, .
\ee
Let us also note that $l \geq 1$, since the solutions with $l = 0$ are pure gauge.
This yields the spectrum of one-particle states in the AdS Fock space \cite{Terashima:2017gmc}. A general solution of the Maxwell field equations in AdS with the boundary conditions \eqref{eq:MagneticBC} can thus be expanded as
\be\label{eq:ModeExpA}
{\cal A}_{\mu}=\sum_{\kappa\in \mathbb{Z}^+}\sum_{l,m} \left[  \frac{1}{\mathcal N^V} a^V_{\kappa,l,m}  {\cal A}^{V,\kappa,l,m}_{\mu}+ \frac{1}{\mathcal N^S} a^S_{\kappa,l,m} {\cal A}^{S,\kappa,l,m}_{\mu}  \right]+\text{c.c.}\, ,
\ee
where ${\cal A}^{V/S,\kappa,l,m}_{\mu}$ are the modes in \eqref{eq:Asols} with their respective quantized positive frequencies, while ``c.c." constains the contributions from the negative frequencies. The normalization constants
\be
\begin{split}
\mathcal N^V &= \sqrt{\frac{4 \Gamma(\kappa + \frac 3 2) \Gamma(\kappa+l+2)}{l(l+1)\pi \Gamma(\kappa+1)\Gamma(\kappa+ l + \frac 3 2)L^2}}\, , \\ 
\mathcal N^S &= \sqrt{\frac{\Gamma(\kappa + \frac 1 2) \Gamma(\kappa+l+1)}{l(l+1)\pi \Gamma(\kappa+1)\Gamma(\kappa+ l + \frac 3 2) L^2}}\, ,
\end{split}
\ee
ensure that the field is unit normalized with respect to the inner product induced by the symplectic potential.
The boundary operator can now be defined simply as
\be
\label{eq:jaVS}
j_{\mu}=\lim_{\rho\rightarrow {\pi \over 2}} \cos^{-1}\rho\, {\cal A}_{\mu}.
\ee
Regarding this formula as an operator equation, the objective is to invert it and obtain the operators $a^{V/S}_{\kappa,l,m}$ in terms of $j_{\mu}$. In the scalar case if scalar fields, the modes are simple spherical harmonics and exponential functions of global time. Here, the math is slightly more complicated due to the appearance of derivatives. Those complications can be bypassed by reconstructing $a^{V/S}_{\kappa,l,m}$ in terms of derivatives of the current operator $j_{\mu}$. This is done by inverting formula \eqref{eq:jaVS} after acting with derivatives with respect to the coordinates at the boundary in both sides. For example, for outgoing modes, the result reads
\be
\begin{split}
a^{V}_{\kappa,l,m} = &{\mathcal N_V \over -i l(l+1)} \frac 1 \pi \int_{0}^{\pi} d\tau'\, \int d^2\Omega'\, 
\epsilon_{\tau'}^{\,\, a b}\nabla_a j^+_b(x') \, Y_l^m(\hat{\Omega}')^*   e^{-i (2+l+2\kappa) \tau'}\, , \\
a^{S}_{\kappa,l,m} =& {\mathcal N_S \over{-i l(l+1)}\omega_S} \frac 1 \pi \int_{0}^{\pi} d\tau'  \int d^2\hat{\Omega}' \,\nabla^{a}j^{+}_a(x') \, Y_l^m(\hat{\Omega}')^*   e^{-i (1+l+2\kappa) \tau'} \, .
\end{split}
\ee
Here, $j^+_{\mu}$ is the positive frequency part of the current and $\epsilon_{a b c}$ is the boundary metric compatible Levi-Civita tensor.The latin indices run over the directions in the $S^2$. The negative frequency part appears in the terms ``c.c." written in formula \eqref{eq:ModeExpA}. Naively we could plug this back in our expression for the bulk gauge field, but a subtlety must be first addressed. As in the scalar case, we will extend the sum over $\kappa$ to all integers and not just the positive ones. This can be done trivially, as the added modes integrate to zero against the positive frequency part of the current $j^+_{\mu}$. We thus conclude
\be
\begin{split}
\label{eq:Akernel}
{\cal A}_{\mu}(x)= \int d^3 x'    \left[ K_{\mu}^{V}(\rho,x,x')\, \epsilon_{\tau'}^{\,\, a b}\nabla_a j^+_b(x') +  K_{\mu}^{S}(\rho,x,x') \,  \nabla^{a}j^+_{a}(x')  \right]\, +\text{h.c}\, .
\end{split}
\ee
where we have defined two different kernels
\be
\begin{split}
K^{V}_{\mu}(x,x')=&\mathcal N_V \frac 1 \pi \sum_{\kappa,l,m } { Y_l^m(\hat{\Omega}')^* \over{-il(l+1)}} {\cal A}^{V,\kappa,l,m}_{\mu}e^{-i\omega_V \tau'}\, , \\
K^{S}_{\mu}(x,x')=&\mathcal N_S \frac 1 \pi \sum_{\kappa,l,m } { Y_l^m(\hat{\Omega}')^* \over{-il(l+1)\omega_S}} {\cal A}^{S,\kappa,l,m}_{\mu} e^{-i\omega_c\tau'} \, .
\end{split}
\ee
One can integrate equation \eqref{eq:Akernel} by parts along the $S^2$ and re-express this results as integrals over the current, instead of its derivatives.

\subsection{Reconstruction with ``electric'' boundary conditions}\label{MaxfieldReconstructionElectric}
We now look for a bulk gauge field with the fall-off behavior
\be\label{eq:ElectricBC}
{\cal V}_{\mu}(\rho,x)\xrightarrow[\rho \to \frac \pi 2]{} \, A_{\mu}(x)\, .
\ee
Here, this defines the normalization of the CFT gauge field two-point function.
The solutions of the Maxwell field equations can still be written as in equation \eqref{eq:Asols},
\be\label{eq:Vsols}
\begin{split}
{\cal V}^{V,l,m}_{\mu} =& L^{-2}\epsilon_{\tau\rho\mu}^{\quad\;\;\nu} e^{\pm i\omega \tau} \cos^{2}\rho \,R_V(\rho) \, \nabla_{\nu} Y_l^m(\hat{\Omega}) \, , \\
{\cal V}^{S,l,m}_{\mu} =& L^{-2}(l(l+1)-\omega^2)\epsilon_{\theta\phi\mu}^{\quad\;\;\nu} \sin^{-1}\theta \,\cot^2\rho\, Y_l^m(\hat{\Omega})\, \nabla_{\nu}(e^{\pm i\omega \tau} R_S(\rho))  \, ,\\
{\cal V}^{G,l,m}_{\mu} =& L^{-2}\nabla_{\mu}\left[ e^{\pm i\omega \tau}Q(\rho) Y_l^m(\hat{\Omega})   \right]   \, ,
\end{split}
\ee
together with the gauge transformation \eqref{eq:gaugeToLorenz} acting on the scalar type solution.
The radial dependence of of the normalizable modes is now exchanged with respect to the magnetic case, so we have
\be
\begin{split}
R_S(\rho)=& \sin^{l+1}\rho\, \cos\rho \, {}_2 F_1\left(  {2+l-\omega\over 2}, {2+l+\omega\over 2};{3\over 2}\Big\vert \cos^2\rho   \right) \, , \\
R_V(\rho)=& \sin^{l+1}\rho\, {}_2 F_1\left(  {1+l-\omega\over 2}, {1+l+\omega\over 2};{1\over 2}\Big\vert \cos^2\rho   \right) \, .
\end{split}
\ee
Demanding regularity of the solutions around the center of AdS$_4$ yields the same quantization of frequencies
\be
\begin{split}
\omega_V=2+l+2\kappa\, , \quad \text{with}\quad \kappa\in \mathbb{Z}^+\, ,\\
\omega_S=1+l+2\kappa\, , \quad \text{with}\quad \kappa\in \mathbb{Z}^+\, .
\end{split}
\ee
The full gauge connection can thus be written as
\be\label{eq:ModeExpV}
{\cal V}_{\mu}=\sum_{\kappa\in \mathbb{Z}^+}\sum_{l,m} \left[  \frac{1}{\mathcal N'_V } v^V_{\kappa,l,m}  {\cal V}^{V,\kappa,l,m}_{\mu}+ \frac{1}{\mathcal N'_S } v^S_{\kappa,l,m} {\cal V}^{S,\kappa,l,m}_{\mu}  \right]+\text{h.c}\, .
\ee
We now want to obtain the coefficients $v^{V/S}_{\kappa,l,m}$ in terms of the boundary operator $b_{\mu}$. Following the same method as above, we find
\be
\begin{split}
v^S_{l,m}=& {\mathcal N'_S \over -l(l+1)} {1\over \pi}\int_{-{\pi \over 2}}^{\pi \over 2}  d\tau'\int d^2\Omega' \, \epsilon_{\tau'}^{\quad ab} \nabla_{a}A^+_{b}(x')e^{-i\omega_S \tau'} Y_l^m(\Omega')^*\, , \\
v^V_{l,m} =&  {\mathcal N'_V  \over  -l(l+1)} {1\over \pi}\int_{-{\pi \over 2}}^{\pi \over 2} d\tau'\int d^2\Omega' \,\nabla_{a} A^{+a}(x')e^{-i\omega_{V}\tau'} Y_l^m(\Omega')^*\, .
\end{split}
\ee
 Much like before, we will sum over all values of $\kappa$ and not just the positive ones. This can be done without trouble as the extra modes integrate to zero against the positive frequency part of $\tilde{j}^{\mu}$. A similar statement can be made for the modes appearing in the ``h.c.'' part of equation \eqref{eq:ModeExpV}, which involve the negative frequencies.  We conclude that the result is
\be
\begin{split}
{\cal V}_{\mu}(x)= \int d^3 x'    \left[ \tilde{K}_{\mu}^{S}(\rho,x;x')\, \epsilon_{\tau'}^{\,\, a b}\nabla_a A^+_b(x') +  \tilde{K}_{\mu}^{V}(\rho,x;x') \,  \nabla^{a} A^+_{a}(x')  \right]\, +\text{h.c}\, .
\end{split}
\ee
where we have defined two different kernels
\be
\begin{split}
\tilde{K}^{S}_{\mu}(\rho,x;x')=& \mathcal N'_S {1\over \pi} \sum_{\kappa,l,m } { Y_l^m(\Omega')^* \over{-l(l+1)}}  {\cal V}^{S,\kappa,l,m}_{\mu} e^{-i\omega_S\tau'} \, , \\
\tilde{K}^{V}_{\mu}(\rho,x;x')=& \mathcal N'_V {1\over \pi} \sum_{\kappa,l,m } { Y_l^m(\Omega')^* \over{-l(l+1)}}   {\cal V}^{V,\kappa,l,m}_{\mu}e^{-i\omega_V \tau'}\, .
\end{split}
\ee

\section{Parameter $\alpha(x)$}\label{app:alpha}
In this appendix we justify the choice of parameter $\alpha(x)$ made in section \ref{sec:WardSoftE}. The logic we follow mirrors the strategy used in flat space when deriving Weinberg soft theorems from the asymptotic symmetries of the QED S-matrix. When only massless particles are present in flat space, ${\cal I}^{\pm}$ can be considered complete Cauchy slices, and the choice of parameter in formula \eqref{eq:varepsilonchoice} yield the desired results. We repeat here this choice.
\be 
\varepsilon(x)={1\over z-z'} \, , \quad \text{and}\quad \varepsilon(x)={1\over \bar{z}-\bar{z}'} \, .
\ee
When discussing massive particles, one must extend this choice to time-like infinity $i^{\pm}$. This is done by realizing that the parameter $\varepsilon(\hat{x})$ is associated to a large gauge transformation, such that
\be
A_{\mu}=\nabla_{\mu}\Lambda\, , \quad \text{with}\quad \Lambda\vert_{{\cal I}^{+}}=\varepsilon(\hat{x})\, .
\ee
Such a large transformation can now be extended into the bulk of Minkowski space by making use of Lorenz gauge
\be
\nabla^{\mu}A_{\mu}=\nabla^2\Lambda=0\, .
\ee
The solution is simply
\be
\Lambda(x)=\int d^2\hat{x}'\,  G(x;\hat{x'})\, \varepsilon(\hat{x}')\, ,
\ee
for a kernel $G(x;\hat{x'})$ obeying the massless Laplace equation and the boundary condition
\be\label{eq:BoundaryG}
\lim_{x\rightarrow {\cal I}^{+}}G(x;\hat{x'}) = \delta^{(2)}(\hat{x},\hat{x}')\, .
\ee
Such a kernel can be written explicitly as follows. A solution of the massless Laplace equation can be expanded in the following set of regular modes
\be
\Lambda_{l,m}(x)=Y_l^m(\hat{x}) F_l\left(  {r\over u} \right)\, , \quad \text{with}\quad F_l(R)=R^l\, {}_2F_1\left(  l,l+1;2l+2|-2R  \right)\, .
\ee
We will thus look for a solution of the form
\be
G(x;\hat{x'})=\sum_{l,m}c_{l,m}\Lambda_{l,m}(x)\, .
\ee
At the null boundary, we have
\be
\lim_{r\rightarrow \infty}G(x;\hat{x'})=\sum_{l,m}c_{l,m}Y_l^m(\hat{x})  {{\Gamma(2+2l)}\over{2^l \Gamma(l+1)\Gamma(2+l)}}\, .
\ee
This is indeed a delta function as in \eqref{eq:BoundaryG} if
\be
c_{l,m}=Y_l^m(\hat{x}')^*  {{2^l \Gamma(l+1)\Gamma(2+l)}\over{\Gamma(2+2l)}}  \, .
\ee
we thus conclude
\be
G(x;\hat{x'})=\sum_{l,m}{{2^l \Gamma(l+1)\Gamma(2+l)}\over{\Gamma(2+2l)}}   Y_l^m(\hat{x}')^* Y_l^m(\hat{x}) F_l\left(  {r\over u} \right)\, .
\ee
Performing the sum explicily yields
 \be
G(x;\hat{x'})=  {1\over 4\pi }{ { t^2-r^2 }  \over {\left( t-r\hat{x}\cdot\hat{x}'  \right)^2 } } \, .
 \ee
This justifies the choice \eqref{eq:MinkepChoice} in the main text.

The calculation in AdS is very similar. In the absence of massive particles, the CFT regions $\tilde{\cal I}^{\pm}$ are the only ones playing a role, and the choice $\alpha(x)=\varepsilon(\hat{x})$ yield the correct result. When massive particles are included in the calculation, one must choose the value of $\alpha(x)$ in the Euclidean half-spheres $\partial{\cal M}_{\pm}$. We define them by noting that $\alpha(x)$ corresponds to a large gauge transformation, such that it can be extended into the bulk using Lorenz gauge. We thus must solve the massless Laplace equation for a parameter $\Lambda(x)$ with boundary conditions
\be
\Lambda(x)\vert_{\tilde{\cal I}^{+}}=\varepsilon(\hat{x})\, .
\ee
We will use the following regular modes obeying the massless Laplace equation
\be
\Lambda_{l,m}(x) =Y_l^m(\hat{x})\, F_l\left(  {{\sin\rho}\over  { \sin\tau-\sin\rho}}  \right)\, , \quad \text{with}\quad F_l(R)=R^l\, {}_2F_1\left(  l,l+1;2l+2|-2R  \right)\, .
\ee
The region $\tilde{\cal I}^+$ corresponds to taking $\rho\rightarrow {\pi \over 2}$ as well as $\tau\rightarrow {\pi \over 2}$, meaning that the argument in $F_l(R)$ becomes large. In such a limit, we have
\be
\Lambda(x)\vert_{\tilde{\cal I}^{+}}=\sum_{l,m}c_{l,m}Y_l^m(\hat{x})  {{\Gamma(2+2l)}\over{2^l \Gamma(l+1)\Gamma(2+l)}}\, .
\ee
We thus conclude that the gauge parameter in the bulk reads
\be
\Lambda(x)=\int d^2\hat{x}'\, \varepsilon(\hat{x}')\, \sum_{l,m} {{2^l \Gamma(l+1)\Gamma(2+l)}\over{\Gamma(2+2l)}}   Y_l^m(\hat{x}')^* Y_l^m(\hat{x}) F_l\left(  {{\sin\rho}\over  { \sin\tau-\sin\rho}} \right)\, .
\ee
performing the sum explicitly yields
\be
\Lambda(x)=\int d^2 \hat{x}' \, {1\over 4\pi}{{\cos^2\rho-\cos^2\tau }\over{\left(   \sin\tau -\sin\rho\, \hat{x}\cdot\hat{x}'  \right)^2}} \varepsilon(\hat{x}')\, .
\ee
The value of this gauge parameter in the Euclidean half spheres is obtained by sending $\rho\rightarrow {\pi \over 2}$, as well as analytically continuing $\tau$ in the imaginary direction. For $\partial{\cal M}_+$, we have $\tau={\pi \over 2}+i\tilde{\tau}$, such that
\be
\Lambda(x)\vert_{\partial{\cal M}_+}=\int d^2 \hat{x}' \, {1\over 4\pi}{{\sinh^2\tilde{\tau} }\over{\left(  \cosh\tilde{\tau} -  \hat{x}\cdot\hat{x}'  \right)^2}} \varepsilon(\hat{x}')\, .
\ee
This concludes the justification for formula \eqref{eq:alphaCFT} in the main text.

\section{AdS Li\'enard-Wiechert potentials}\label{app:LWAdS}
In this appendix we compute the Li\'enard-Wiechert potential associated to a particle propagating in Anti-de Sitter space-time. Using the linearity of electro-magnetism in the region of asymptotic decoupling, one can build solutions involving several particles by adding the solutions of the form presented here. 

Li\'enard-Wiechert gauge fields are classical solutions to the heterogeneous Maxwell equation
\be
\nabla_{\mu}F^{\mu\nu} = J^{\mu}\, .
\ee
The current for a single charged particle reads explicitly
\be
J^{\mu}(Y)=q \int ds\,  {{\partial X^{\nu}(s)}\over{\partial s}} \, \delta^{(4)}\left(Y-X(s)\right)\, .
\ee
We will first analyze massive particles. The strategy we will follow will be to find the solution for a stationary particle located at the origin, and then use the AdS isometries to obtain the solution for a particle following a generic time-like geodesic. The solution for the stationary case in global coordinates reads 
\be
{\cal A}^{(\text{C})}_{\mu}= q \cot \rho  \, \delta^{\tau}_{\mu}\, ,
\ee
where the label $(\text{C})$ is a reminder that we are chosing a particular homogeneous (radiative) solution, namely that of no radiation if the particle is at rest. The solution written above diverges at the location of the charge, $\rho=0$. We now act with an AdS isometry on this solution, to obtain the gauge field associated to a particle moving along a generic time-like geodesic.  

Isometries of AdS can be easily described in embedding space, with line element
\be
ds^2=-dX_1^2-dX_2^2+dX^2_3+dX^2_4+dX^2_5\, .
\ee
The  embedding of AdS in such space reads
\be\label{eq:Embedding}
\begin{split}
X_{1} &= L {{\cos\tau}\over{\cos\rho}}\, , \quad
X_{2} = L {{\sin\tau}\over{\cos\rho}}\, , \\
X_{3} &= L \tan\rho \sin\theta \cos\phi \, ,\quad
X_{4} = L \tan\rho \sin\theta \sin\phi \, ,\quad
X_{5} = L \tan\rho \cos\theta \, .
\end{split}
\ee
The $SO(2,4)$ isometries of AdS are then translations, boosts, and rotations in embedding space. We are interested only in transformations that keep the particle at the origin at $\tau=0$, such that the particle remains inside the scattering region as the flat limit is taken. We further require a transformation that makes the new geodesic hit the point $\tau=\pi / 2$ with $\cos\rho=m/ \omega$ at an angle $\hat{p}$. Such a geodesic will hit the complexified boundary $\partial_+{\cal M}$ at the point \eqref{eq:Imtau}, which means that this geodesic is the one followed by particles creating the desired scattering states in Minkowski space-time. Such geodesics can be obtained by transforming embedding coordinates with a boost along $X_2$ and a rotation. Explicitly,
\be
X^{\prime I}=\Lambda^{I}_{J} X^J\, , \quad \text{with} \quad \Lambda^{I}_J=R^I_K B^K_J\, ,
\ee
with rotation and boost matrices given by
\be
R=\left(     \begin{matrix}  1&0&0&0&0\\  0&1&0&0&0  \\ 0&0&-\sin\theta_p\cos\phi_p  &  \cos\theta_p\cos\phi_p & \sin\phi_p \\  0&0&-\sin\theta_p\sin\phi_p & \cos\theta_p\sin\phi_p & -\cos\phi_p   \\  0&0& -\cos\theta_p & -\sin\theta_p &0 \end{matrix}    \right)
\, , \quad 
B=\left(     \begin{matrix}  1&0&0&0&0\\  0&{\omega \over m}&  -{\vert \vec{p}\vert\over m} &0&0  \\ 0&-{\vert \vec{p}\vert \over m}&{m\over \omega} &  0 &0 \\  0&0&0 & 1 & 0  \\  0&0& 0& 0 &1 \end{matrix}    \right)\, .
\ee
The embedding space vector $X'$ defines new coordinates $\rho',\tau',\hat{x}'$ in the form of an embedding like equation \eqref{eq:Embedding}. Explicitly, the new coordinates are related to the old ones as follows
\be
\begin{split}
\tan\tau&={{\omega\sin\tau'-\vert\vec{p}\vert \sin\rho' \Omega'\cdot \hat{p}}\over{m \cos\tau'}}\, , \\
\cos\rho&={{\cos\rho'}\over{\cos\tau'}}  {1\over{\sqrt{1-\left( {{   \omega\sin\tau' -\vert \vec{p}\vert \sin\rho' \hat{x}'\cdot\hat{p}}\over{m\cos\tau'}}  \right)^2}}}\, ,\\
\cos\theta &= -{{\sin\rho'}\over{\cos\tau'}} {{   \partial_{\phi_p}(\hat{p}\cdot\hat{x}')  }\over{  \sin\theta' }}  {1\over{\sqrt{   1-{{\cos^2\rho'}\over{\cos^2\tau'}} - \left( {{   \omega\sin\tau' -\vert \vec{p}\vert \sin\rho' \hat{x}'\cdot\hat{p}}\over{m\cos\tau'}}  \right)^2   }}}\, , \\
\tan\phi&=  {{ m\sin\rho'   }\over{   \vert \vec{p}\vert  \sin\tau' -\omega \sin\rho' \hat{p}\cdot\hat{x}'  }}  \partial_{\theta_p}(\hat{p}\cdot\hat{x}' )   \, .
\end{split}
\ee
With these expressions at hand, we are now ready to compute
\be
{\cal A}_{\mu'}^{(\text{C})}(x')={{\partial x^{\mu}}\over{\partial x^{\mu'}}} {\cal A}_{\mu}^{(\text{C})}(x)\, .
\ee
The result for the gauge field is generally complicated. However, expressions simplify when approaching the boundary, which is where we want to evaluate the Li\'enard-Wiechert fields. For example, at $\rho'={\pi\over 2}$ and $\tau'={\pi \over 2}$, we have
\be\label{eq:ResultC1}
\langle j_{\tau'} \left(  \tau'={\pi \over 2}, \hat{x}'     \right)\rangle = \lim_{\rho'\rightarrow{\pi \over 2},\tau'\rightarrow{\pi \over 2}} {1\over \cos\rho'} {\cal A}_{\tau'} = q{{m^2}\over{\left(  \omega- p\cdot \hat{x}'      \right)^2}}\, .
\ee

This result can be used to make the argument of section \ref{sec:WardSoftE}. There, we used that
\be\label{eq:argument}
\int_{\partial \mathcal M_+} d^3 x \, \partial_{\mu}\alpha(x) \langle j^{\mu}\rangle =0 \, ,
\ee
where, as a reminder, the parameter $\alpha(x)$ is chosen to be 
\be\label{eq:alphaCFTRepeat}
\alpha(x)=\lim_{\rho\rightarrow{\pi \over 2}}\int d^2 \hat{x}' \, {1\over 4\pi}{{\cos^2\rho-\cos^2\tau }\over{\left(   \sin\tau -\sin\rho\, \hat{x}\cdot\hat{x}'  \right)^2}} \varepsilon(\hat{x}')\, .
\ee
To see this, we first integrate by parts
\be
\int_{\partial {\cal M}_+} d^3 x \, \partial_{\mu}\alpha(x) \langle j^{\mu}\rangle =\int_{\tau={\pi \over 2}} d^2\hat{x} \, \varepsilon(\hat{x})\,  \langle j_{\tau} \rangle - \int_{\partial {\cal M}_+}  d^3 x \, \alpha(x) \partial_{\mu}\langle j^{\mu}\rangle \, .
\ee
In the first term we have used the fact that the parameter $\alpha(x)$ becomes $\varepsilon(\hat{x})$ in the CFT region $\tilde{\cal I}^+$. This term can be written explicitly using \eqref{eq:ResultC1}, which results in
\be\label{eq:AUX1}
\int_{\tau={\pi \over 2}} d^2\hat{x}\, \varepsilon(\hat{x})\,  \langle j_{\tau} \rangle= q \int d^2\hat{x} \varepsilon(\hat{x}) {{m^2}\over{\left(  \omega- p\cdot \hat{x}'      \right)^2}}\, .
\ee

In the second term, we can invoke the Ward identity to obtain a delta function for the insertion of an operator at $\hat{x}=\hat{p}$ and $\tau={\pi \over 2}+i\log\sqrt{{\omega+m}\over{\omega-m}}$. The value of $\alpha(x)$ at that location is simply
\be\label{eq:AUX2}
 \int_{\partial {\cal M}_+}  d^3 x \, \alpha(x) \partial_{\mu}\langle j^{\mu}\rangle = q \int d^2 \hat{x}' \, \varepsilon(\hat{x}') {{m^2}\over{\left(  \omega- p\cdot \hat{x}'      \right)^2}}\, .
\ee
Combining \eqref{eq:AUX2} and \eqref{eq:AUX1} we immediately arrive at the desired result \eqref{eq:argument}. Note that the calculations here involve a single particle in AdS and a single operator insertion in the CFT. However, the bulk calculations can be easily generalized to include more particles due to the linearity of electro-magnetism in the region of asymptotic decoupling.  

In section \ref{sec:WardSoftE}, we also argued that the Coulombic current does not contribute to the second line of \eqref{eq:ArgumentsHolographic}. This can also be checked with the explicit form of the Li\'enard-Wiechert potentials calculated here, which yield vanishing expectation values $\langle j_{z}\rangle$ and $\langle j_{\bar{z}}\rangle$ at $\tilde{\cal I}^+$.

A similar argument was also made for the case of electric boundary conditions in section \ref{sec:MST}. In that case, a topological current was constructed, and the relation of interest is
\be\label{eq:argumentM}
\int_{\partial {\cal M}_+} d^3 x \, \partial_{\mu}\alpha(x) \langle (*f)^{\mu}\rangle =0 \, , \quad \text{with}\quad f=dA \, .
\ee
This can be proven following a similar logic as the one above. We start with a stationary magnetic monopole in anti-de Sitter space. The Li\'enard-Wiechert field associated to such object reads
\be
{\cal V}^{(\text{C})}_{\mu}=g \cos \theta  \, \delta^{\phi}_{\mu}\, .
\ee
One can then act on this result by an isometry of AdS such that the geodesic followed by the particle hits the point of the CFT where an operator must be inserted in order to create a scattering state. A calculation similar to the one above results in
\be
\langle (*f)_{\tau'} \left(  \tau'={\pi \over 2}, \hat{x}'     \right)\rangle = \lim_{\rho'\rightarrow{\pi \over 2},\tau'\rightarrow{\pi \over 2}}  \epsilon_{\tau'}^{\quad a'b'} \partial_{a'}{\cal V}_{b'} = g{{m^2}\over{\left(  \omega- p\cdot \hat{x}'      \right)^2}}\, .
\ee
This result mirrors the one concerning electrical charges in formula \eqref{eq:ResultC1}. The rest of the proof leading to \eqref{eq:argumentM} follows verbatim.

\section{Sample calculations}
\label{app:sample_calc}
In this section, we will calculate two simple quantities. This mostly serves as a sanity check and to illustrate calculation techniques. Throughout the paper we have chosen normalizations such that a canonically normalized CFT scalar two-point function gives a correctly normalized S-matrix element. Hence, there is no need to fix the normalization.
\subsection{One-to-one scattering}
As an example, we will compute the one-to-one scattering amplitude of a massless scalar field.
\begin{align}
\begin{split}
    \label{eq:two_pt}
    & \qquad \langle{a_{\text{out},q}a^\dagger_{\text{in},p}}\rangle = \\
 &(-1)\sqrt{\frac{24^2}{L^8 \omega_p^5\omega_q^5}}\iint_{-\frac \pi 2 L}^{\frac\pi 2 L} dt dt' e^{i \omega_q L t - i \omega_p L t'} \langle \mathcal O^-\left( \frac{t'} L + \frac \pi 2, \hat q\right) \mathcal O^+\left(\frac{t} L - \frac \pi 2, - \hat p\right) \rangle.
\end{split}
\end{align}
It is now helpful to use the spectral decomposition of the operators. For a free theory we have that
\begin{align}
    \mathcal O^\pm(\tau, \Omega) = \sum_{\kappa,l,m} \sqrt{\frac{4\pi}{3} \frac{\Gamma(\kappa + l + 3)\Gamma(\kappa + \frac 5 2)}{\Gamma(\kappa+1) \Gamma(\kappa + l + \frac 3 2)}} e^{\pm i \omega_{\kappa,l} \tau} Y^m_l(\Omega) a^{\pm}_{\kappa,l,m}.
\end{align}
The creation/annihilation operators $a^{\pm}_{\kappa,l,m}$ act on the CFT vacuum and result in Kronecker-deltas which set the $\kappa, l, m$ modes to be the same for both operators.
Plugging this into equation \eqref{eq:two_pt} we obtain after performing the $t$ integrals,
\begin{align}
\begin{split}
\langle{a_{\text{out},q}a^\dagger_{\text{in},p}}\rangle  ={}& \frac{-16}{L^4 \omega_p^3} \frac{(2 \pi)^3}{\omega_p^2} \delta(\omega_p - \omega_q)  \\
    & \cdot \sum_{\kappa,l,m} Y^m_l(\hat q) Y^{*m}_l(- \hat p) e^{i \omega_p \pi L} \delta\left(\omega_p - \left(\frac{2 \kappa + l + 3} L\right)\right) \frac{\Gamma(\kappa+l+3)\Gamma(\kappa+\frac 5 2)}{\Gamma(\kappa+1)\Gamma(\kappa+l+\frac 3 2)}.
\end{split}
\end{align}
The phase factor becomes $e^{i \omega_p \pi L} = (-1)^{l + 3}$. We can turn the sum over $\kappa$ into an integral over $x = \frac{2 \kappa} L$. The integration can then be performed using the delta function. This results in
\begin{align}
     \langle{a_{\text{out},q}a^\dagger_{\text{in},p}}\rangle  = \left(\frac{2}{L \omega_p}\right)^3 \frac{(2 \pi)^3}{\omega_p^2} \delta(\omega_p - \omega_q) \sum_{l,m} Y^m_l(\hat q) Y^{*m}_l(\hat p) \frac{\Gamma(\frac{\omega L + l + 3}{2})\Gamma(\frac{\omega L - l + 2}{2})}{\Gamma(\frac{\omega L - l -1}{2})\Gamma(\frac{\omega L + l}{2})}.
\end{align}
In the large $L$ limit, the gamma functions become independent of the angular momentum quantum number $l$. The sum over spherical harmonics can be performed and the  Gamma functions exactly cancel the first factor. Finally, we obtain 
\begin{align}
    \langle{a_{\text{out},q}a^\dagger_{\text{in},p}}\rangle = \frac{(2 \pi)^3}{\omega_p^2} \delta(\omega_p - \omega_q) \delta^{(2)}(\hat p  - \hat q) = (2 \pi)^3 \delta^{(3)}(\vec p - \vec q).
\end{align}

\subsection{One particle-state normalization}
Another aspect of our proposal is that the in- and out- Fock space is constructed using positive and negative frequencies of operators located at $\mp \frac \pi 2$. We should thus be able to calculate inner products between states in any of the asymptotic Fock spaces. For definiteness, we take the out-Fock space, which is obtained by acting with operators $a^\dagger_{\text{out},p}$ on the vacuum. The vacuum is defined by the condition that it is annihilated by all $a_{\text{out},p}$. Here, we will show that the one-particle state is canonically normalized, that is, we want to calculate
\begin{align}
\begin{split}
   & \langle{a_{\text{out},q}a^\dagger_{\text{out},p}}\rangle = \\
 &\qquad  \sqrt{\frac{24^2}{L^8 \omega_p^5\omega_q^5}}\iint_{-\frac \pi 2 L}^{\frac\pi 2 L} dt dt' e^{i \omega_q L t - i \omega_p L t'} \langle \mathcal O^-\left( \frac{t'} L + \frac \pi 2, \hat q\right) \mathcal O^+\left(\frac{t} L + \frac \pi 2, - \hat p\right) \rangle.
\end{split}
\end{align}
This time, after performing the time-integrals, we obtain
\begin{align}
\begin{split}
\langle{a_{\text{out},q}a^\dagger_{\text{out},p}}\rangle  ={}& \frac{16}{L^4 \omega_p^3} \frac{(2 \pi)^3}{\omega_p^2} \delta(\omega_p - \omega_q)  \\
    & \sum_{\kappa,l,m} Y^m_l(\hat q) Y^{*m}_l(\hat p) e^{i \omega_p \pi L} \delta\left(\omega_p - \left(\frac{2 \kappa + l + 3} L\right)\right) \frac{\Gamma(\kappa+l+3)\Gamma(\kappa+\frac 5 2)}{\Gamma(\kappa+1)\Gamma(\kappa+l+\frac 3 2)}.
\end{split}
\end{align}
Note that in comparison with the S-matrix element calculated above, the overall sign and the argument of the spherical harmonics differ. In addition, the phase factor evaluates to $e^{i \omega_p \pi L} = 1$. The rest of the calculation is exactly the same as above and the result reads
\begin{align}
    \langle{a_{\text{out},q}a^\dagger_{\text{out},p}}\rangle = \frac{(2 \pi)^3}{\omega_p^2} \delta(\omega_p - \omega_q) \delta^{(2)}(\hat p  - \hat q) = (2 \pi)^3 \delta^{(3)}(\vec p - \vec q).
\end{align}

\bibliographystyle{JHEP}
\bibliography{refs}
\end{document}